\documentclass[usenatbib,preprint]{mn2e}
\usepackage{graphicx}
\usepackage {natbib,aas_macros}
\usepackage {mediabb}
\usepackage {bm}
\usepackage {amsmath,amssymb}

\newcommand{\msun}{\thinspace M_\odot} 
\newcommand{\gcm}{~{\rm g~cm}^{-3} }

\newcommand{\magB}{\mathbf{B}}
\newcommand{\eleE}{\mathbf{E}}
\newcommand{\cul}{\mathbf{J}}
\newcommand{\vel}{\mathbf{v}}

\newcommand{\cms}{~{\rm cm} ~{\rm s}^{-1} } 
 
\newcommand{\ms}{~{\rm m} ~{\rm s}^{-1} }

\newcommand{\mum}{{\rm \mu} {\rm m} }

\newcommand{\acc}{\mathbf{a_0}}
\newcommand{\dt}{\Delta t}
\newcommand{\dv}{\Delta \vel}
\newcommand{\expm}{{\rm expm1}}
\newcommand{\tstop}{t_{\rm stop}}
\newcommand{\fgp}{\mathbf{f}_{g, p}}
\newcommand{\fgem}{\mathbf{f}_{g, em} }
\newcommand{\fdem}{\mathbf{f}_{d, em}}
\newcommand{\fge}{\mathbf{f}_{g, e} }
\newcommand{\fgm}{\mathbf{f}_{g, m}}

\title{Conditions for justifying single-fluid approximation for charged and neutral dust fluids
  and a smoothed particle magnetohydrodynamics method for dust-gas mixture}

\author[Tsukamoto et al]{
Y. Tsukamoto$^{1}$, M. N. Machida$^{2}$, and  S. Inutsuka$^{3}$ \\
$^1$Graduate Schools of Science and Engineering, Kagoshima University, Kagoshima, Japan  \\
$^2$Department of Earth and Planetary Sciences, Kyushu University, Fukuoka, Japan \\
$^3$Department of Physics, Nagoya University, Aichi, Japan  \\
}

\begin{document}
\maketitle

\begin{abstract}
  We describe a numerical scheme for magnetohydrodynamics
  simulations of dust-gas mixture by extending smoothed particle magnetohydrodynamics.
  We employ the single-species particle approach to describe dust-gas mixture
  with several modifications from the previous studies.
  We assume that the charged and neutral dusts can be treated as single-fluid
  and the electro-magnetic force acts on the gas and  that on the charged dust is negligible.
  The validity of these assumption in the context of protostar formation is not obvious and is extensively evaluated.
  By investigating the electromagnetic force and electric current with terminal velocity approximation,
    it is found that as the dust size increases, the contribution of dust to them becomes smaller and negligible.
  We conclude that our assumptions of the electro-magnetic force on the dusts is negligible are valid
  for the dust size with $a_d \gtrsim 10 \mum$.
  On the other hand, they do not produce the numerical artifact for the dust $a_d \lesssim 10 \mum$ in envelope and disk where
  the perfect coupling between gas and dusts realizes.   However, we also found that our assumptions may break down in outflow
  (or under environment with  very strong magnetic field and low density) for the dust $a_d \lesssim 10 \mum$.
  We conclude that our assumptions are valid in almost all cases where macroscopic dust dynamics
  is important in the context of protostar formation.
  We conduct numerical tests of dusty wave, dusty magnetohydrodynamics shock,
  and gravitational collapse of magnetized cloud core with our simulation code.
  The results show that our numerical scheme well reproduces the dust dynamics in the magnetized medium.
\end{abstract}

\section{Introduction}
\label{intro}
Dust dynamics, as well as  dust growth during protostar formation has drawn attention in recent years.
Dusts are the fundamental building block of planetesimals and planets
\citep[][]{1973ApJ...183.1051G,1985prpl.conf.1100H,2012ApJ...752..106O}.
Dusts affect the ionized state of molecular clouds and young stellar objects (YSOs)
\citep{1990MNRAS.243..103U,1991ApJ...368..181N,2002ApJ...573..199N,2012A&A...541A..35D,2016A&A...592A..18M,2018MNRAS.478.2723Z},
determines the degree of coupling between gas and magnetic field, and has significant influence on the evolution of YSOs
\citep[e.g.,][]{1991ApJ...368..181N,2012A&A...541A..35D,2018MNRAS.473.4868Z,2020ApJ...896..158T}.
The dust thermal emission is important tool for the observations of molecular clouds and YSOs.

Observations of molecular cloud cores and YSOs have suggested possible dust growth
\citep[e.g.,][]{1991ApJ...381..250B,2002ApJ...581..357K,2007ApJ...659..479J,
2009ApJ...696..841K,2010A&A...521A..66R,2010A&A...512A..15R,2010Sci...329.1622P,2010A&A...511A...9S,
2014A&A...567A..32M}.
Surprisingly,  dust growth was found to occur even  in relatively low-density regions.
Observations detected infrared emission from molecular cloud cores, which
is interpreted as scattered light by dusts with sizes of $\gtrsim 1 \mum$,
suggesting the existence of relatively large dusts in molecular cloud cores
\citep{2010Sci...329.1622P,2010A&A...511A...9S}.
Observations of the envelopes in Class 0/I  YSOs show low
opacity-spectral indices, $\beta_s$ at (sub-)mm wavelengths.
These low $\beta_s$ may be an indication for the presence of (sub-)mm size dusts
\citep[][]{2009ApJ...696..841K,2014A&A...567A..32M}.

From a theoretical point of view, however,
the dust growth in the molecular cloud core or
envelope to several $\mum$ to (sub-)mm sizes is very difficult to explain because
the densities in cloud cores and envelopes are low \citep[e.g.,][]{2009A&A...502..845O, 2013MNRAS.434L..70H}.
For example, \citet{2013MNRAS.434L..70H} showed that the formation
of micron-sized grain in the dense cloud core should take several times as long as the free-fall time.
Then the formation of (sub-)mm dust in the envelope should take  $\sim 10$
times the free-fall time even in the optimistic
coagulation conditions.
As such, the observational evidence for the existence of dust of $\sim 1 \mum$ in cloud core or $\sim 1 {\rm mm}$ in envelope
is puzzling for the current dust-growth theory.

The theoretical studies of dust growth in the cloud core or envelope
have been mainly based with the one-zone approximation
in which  the complicated gas dynamics observed in real YSOs
such as disk formation and outflow launching from disk are not considered.
Thus, studies of the interplay of these more realistic processes and dust dynamics
have good potential to bring new insights about dust growth and dust dynamics
in the cloud core and YSOs.

Several attempts have been reported to reveal the dust dynamics
with multi-dimensional simulations.
\citet{2017MNRAS.465.1089B} investigated the dust
dynamics during gravitational collapse of the unmagnetized cloud core and
showed that the dust-to-gas
mass ratio is enhanced in the central region if the dust already grows to $\gtrsim 100 \mum$ in size in cloud cores.
\citet{2018A&A...614A..98V, 2019A&A...627A.154V} have studied
the gas and dust dynamics,  including dust growth with two-dimensional simulation and
reported a change of the dust-to-gas ratio in the disk.
However, these studies were hydrodynamics simulations and the magnetic field was ignored.

One difficulty in investigating the dust dynamics in the magnetized medium is
the handling of the charged dusts. In molecular clouds and YSOs,
a large portion of the dusts are charged absorbing electrons and ions and are subject to
electromagnetic force.
Because the charge neutrality is achieved as the dust-gas mixture,
it is not achieved in the gas phase or dust phase alone. Hence
the electric force on gas and dust is, in principle, not negligible.
Furthermore, \citet{2002ApJ...573..199N} pointed out that, with the dust size of the interstellar medium,
the dusts should be the main receivers of the electromagnetic force,
and friction force between the neutral gas and charged dusts should be
much larger than that between the neutral gas and ions.

Treating charged-dust and charged-gas
as a multi-fluid is numerically very difficult and expensive
as treating ion and electron as a two-fluid.
Furthermore, unlike fully ionized plasma,
the abundances of the ions, electrons, and charged dusts vary as the gas density changes through
chemical reactions in the molecular clouds and YSOs.
This fact adds more difficulty to the numerical simulation.
Therefore, appropriate approximation for the electro-magnetic force
on the gas and the dust phases is required.

It is expected that the electro-magnetic force becomes less significant as dust
size increases, whereas the dust advection (or macroscopic dust dynamics)
becomes significant only when the dust has grown sufficiently large.
Given that these two effects roughly counteract each other in practice,
an approximation with good properties is valid in the broad
parameter range of the density and magnetic field for cloud core and YSOs
as we demonstrate in \S 2.

Very recently, \citet{2020A&A...641A.112L} investigated the dust dynamics of the magnetized cloud core.
They considered the neutral dust only to avoid the technical difficulty described above.
Although they argued that
the electro-magnetic force on the dust is negligible on
the basis of the argument with the the Hall parameter
in their discussion and justified their strategy,
we consider that their discussion is inadequate because the Hall parameter only indicates
whether the magnetic
force or frictional force is balanced with the electric force \citep{2007Ap&SS.311...35W} and
does not guarantee that the electro-magnetic force on the dust is negligible.
Therefore, we believe that more detailed study is required to reveal the condition in which the
electro-magnetic force on the dusts is negligible.

We here report our development of a numerical
scheme for dust-gas mixture in the magnetized medium for
smoothed particle magnetohydrodynamics
assuming that the charged and neutral dusts can be treated as single-fluid
and the electro-magnetic force acts on the gas and that on the dust is negligible.
The validity of these assumption is not obvious.
Thus, we extensively evaluate them in the context of protostar formation.
We employ the approach proposed by \citet{2014MNRAS.440.2136L} and \citet{2014MNRAS.440.2147L}
in which the dust-gas mixture is represented by single-species SPH particles which is advantageous
for the tightly coupled dust-gas mixture.
In this paper we focus on  approximations, code implementations, and  numerical tests. Our
in-depth and scientific results will be presented in the future papers.

This paper is organized as follows.
In \S 2, we describe the governing equations and approximations employed in our numerical scheme.
We also investigate the validity of the approximations.
In \S 3, the discretization of our  numerical scheme.
Then in \S 4, we describe the numerical tests.
Finally, our results are summarized in \S 5.

\section{Governing equations and approximations}
\label{governing_equation}
\subsection{the equation of continuity and the momentum equation}
In this section, we derive the  equation of continuity and the momentum equation
employed in our numerical scheme  starting from  the single-fluid equations for gas,
\begin{eqnarray}
  \label{gas_single_fluid_first}
  \frac{D_g \rho_g}{D t}&=&- \rho_g \nabla \cdot \vel_g, \\
  \frac{D_g \vel_g }{D t}  &=& \frac{(\vel_d-\vel_g)}{\tstop} + \frac{1}{\rho_g} \{ -\nabla P_g+ (\tau_g \eleE+ \frac{\cul_g \times \magB}{c}) \} + \mathbf{f},
  \label{gas_single_fluid_last}
\end{eqnarray}
and multi-fluid equations for dust in which charged and neutral dusts are considered separately,
\begin{eqnarray}
  \label{dust_multi_fluid_first}
  \frac{D_{d,Z} n_{d,Z}}{D t} &=&- n_{d,Z} \nabla \cdot \vel_{d, Z}, \\
  \frac{D_{d,Z} \vel_{d,Z} }{D t}  &=& -\frac{\vel_{d,Z}-\vel_g}{\tstop} + \frac{q_{Z}}{m_{d}} \{(\eleE+  \frac{\vel_{d, Z} \times \magB}{c}) \} + \mathbf{f},
  \label{dust_multi_fluid_last}
\end{eqnarray}
where subscript $Z$ denotes the charge number of the dusts ($Z=0,\pm 1,\pm 2,\cdots$).
In this paper, we consider monosized dusts.
We define $D_{[g, (d,Z)]}/Dt=\partial/\partial t + \vel_{[g, (d,Z)]} \cdot \nabla$,
where a pair of square brackets ($[]$) indicates any one of the symbols inside
("$g$" and "$(d, Z)$" for gas and dust, respectively) applied throughout the equation.
$\rho_{g}$ is  gas  density,
$n_{d, Z}$  is  dust number density,
$m_d$ is the mass of the dust,
$q_{Z}$  is  dust charge,
$\vel_{[g, (d,Z)]}$ are gas and dust velocity,
$\vel_d$ is the barycentric velocity of the dusts.
$t_{\rm stop}$ is the stopping time, 
$P_g$ is the gas pressure,  
$\magB$ is the magnetic field,
$\eleE$ is the electric field,
$\tau_g=\sum_{j=g} q_j n_j$, $\cul_{g}=\sum_{j=g} q_j n_j \vel_j$
are the charge density and electric current of gas phase,
where $\sum_{j=g}$  mean the summations of the gas-phase species,
and $\mathbf{f}$ is the specific force which acts both on gas and dust such as gravity.
Here we assume $\tstop$ does not depends on the charge \citep[see e.g., ][]{2008A&A...484...17P}.

The reason why we start from the multi-fluid equations for dusts is as follows.
In the molecular cloud core or around the protostars,
the majority of the dusts can be negatively charged.
Thus, unlike gas phase in which most of the component is neutral,
the charged dusts has non-negligible abundances, in particular,
in low density region or in the case that the dust size is large.
The charged dusts receive the electro-magnetic force {depending on their charge}
and the neutral dusts do not, hence
the charged dusts and neutral dusts may have different velocities.
Therefore, in principle, we should treat them separately.

However, as  we will discuss in the next subsection, the relative velocities among the charged
and neutral dusts are very small in almost all situation where the dust dynamics is important
in the molecular cloud core and YSOs and, surprisingly, single-fluid approximation for dusts is valid (see
\S \ref{rel_vel_dusts} and figure \ref{relative_velocity_among_dust}).

By summing up about the subscript $Z$ of equations (\ref{dust_multi_fluid_first}) and
(\ref{dust_multi_fluid_last}), and neglecting the second order terms of relative velocity
among the dusts,
we obtain single-fluid equations for dusts,
\begin{eqnarray}
  \label{dust_single_fluid_first}
  \frac{D_d \rho_d}{D t}&=&- \rho_d \nabla \cdot \vel_d \\
  \frac{D_d \vel_d }{D t}  &=& - \frac{(\vel_d-\vel_g)}{\tstop} + \frac{1}{\rho_d} \{(\tau_d \eleE+ \frac{\cul_d \times \magB}{c}) \} + \mathbf{f}
  \label{dust_single_fluid_last}
\end{eqnarray}
where
$D_{d}/Dt=\partial/\partial t + \vel_{d} \cdot \nabla$.
$\rho_d=\sum_Z m_d n_{d,Z}$ is the mass density of the dusts
$\tau_d=\sum_Z q_{Z} n_{d,Z}$ is the charge density of the dusts
$\cul_d=\sum_Z q_{Z} n_{d,Z} \vel_{d, Z}$ is the electric current of the dusts

Following \citet{2014MNRAS.440.2136L} and \citet{2014MNRAS.440.2147L},
equations (\ref{gas_single_fluid_first}),  (\ref{gas_single_fluid_last}),
(\ref{dust_single_fluid_first}), and  (\ref{dust_single_fluid_last}) can be rewritten as 
\begin{eqnarray}
  \label{single_fluid_continum}
  \frac{D \rho}{D t}  &=& - \rho \nabla \cdot \vel, \\
  \label{single_fluid_eps}
  \frac{D \epsilon}{D t} &=& -\frac{1}{\rho}\cdot \{\epsilon(1-\epsilon)\rho \Delta \vel \}, \\
  \label{single_fluid_motion1}
  \frac{D \vel }{D t} &=& \frac{1}{\rho} \{ -\nabla P_g+ (\tau \eleE+ \frac{\cul \times \magB}{c}) \} \nonumber \\
  &-&\frac{1}{\rho} \nabla \cdot \{ \epsilon (1-\epsilon)\rho \dv \dv\} + \mathbf{f}  \\
  \label{single_fluid_motion2}
  &\sim& \frac{1}{\rho} \{ -\nabla P_g+ ( \frac{\cul \times \magB}{c}) \} \nonumber \\ 
  &-& \frac{1}{\rho} \nabla \cdot \{ \epsilon (1-\epsilon)\rho \dv \dv\} + \mathbf{f}, \\ 
  \label{single_fluid_dv2}
  \frac{D \dv }{D t} &=&-\frac{\dv}{\tstop}+\{ \frac{1}{\rho_g}[-\nabla P_g+ (\tau_g \eleE+ \frac{\cul_g \times \magB}{c})] \nonumber \\
  &-& \frac{1}{\rho_d}(\tau_d \eleE+ \frac{\cul_d \times \magB}{c}) \} \nonumber \nonumber \\
  &-& (\dv \cdot \nabla)\vel+\frac{1}{2} \nabla \{(2\epsilon-1)\dv^2\}  \nonumber \\
  &\equiv&-\frac{\dv}{\tstop}+(\fgp+\fgem -\fdem) \nonumber \\ 
  &-& (\dv \cdot \nabla)\vel+\frac{1}{2} \nabla \{(2\epsilon-1)\dv^2\},
\end{eqnarray}
where
$\rho=\rho_g+\rho_d$ is the total density,
$\epsilon=\rho_d/\rho$,
$\vel=(\rho_g \vel_g+\rho_d \vel_d)/(\rho_g+\rho_d)$ is barycentric velocity of dust gas mixture,
$\dv=(\vel_d-\vel_g)$ is the velocity difference between gas and dust,
$\tau=\tau_g+\tau_d$ is the total charge density,
$\cul=\cul_g+\cul_d$ is the total electric current,
$\fgp=-\frac{\nabla P_g}{\rho_g}$ is the pressure gradient force,
$\fgem=\frac{1}{\rho_g}(\tau_g \eleE+ \frac{\cul_g \times \magB}{c})$ is electro-magnetic force on the gas phase,
$\fdem=\frac{1}{\rho_d}(\tau_d \eleE+ \frac{\cul_d \times \magB}{c})$ is electro-magnetic force on the dust phase,
and $D/Dt=\partial/\partial t + \vel \cdot \nabla$.
In the reduction from equations (\ref{single_fluid_motion1})
to (\ref{single_fluid_motion2}), we use the relation
$\tau \eleE \ll \frac{\cul \times B}{c}$
which is deduced from  Maxwell's equations,
using $O(\tau \eleE)/O(\cul \times B/c ) = O(\eleE^2) /O(\magB^2) = O(v^2/c^2)\ll 1 $.

Equation (\ref{single_fluid_dv2}) corresponds to the generalized Ohm's law if the
velocity difference between the ionized gas and charged dusts determines the electric current.
However, as we will see below, the contribution of the dusts to the electric current is negligible
in many situations in the evolution of molecular cloud cores and YSOs.
Hence, we keep equation (\ref{single_fluid_dv2}) in the above-described form.

The difficulty in solving equations (\ref{single_fluid_continum}) to  (\ref{single_fluid_dv2}) lies in equation (\ref{single_fluid_dv2}).
To solve equation (\ref{single_fluid_dv2}),
we need the information of  $\tau_g$, $\cul_g$, $\tau_d$, $\cul_d$, and $\eleE$ which
are determined from the ionization state of and relative velocity
between the ions, electrons, and charged dusts.
The ionization state varies with the density
and temperature even when the advection
of dust is ignored \citep[e.g.,][]{1990MNRAS.243..103U,2002ApJ...573..199N}.
Thus, treating these processes in three dimensional simulation is very difficult task.
%

However, the sub-micron dusts in the collapsing cloud core
can be assumed to be macroscopically completely coupled with the gas
regardless of their charges and have been treated as such in a number of studies
\citep{1990MNRAS.243..103U,2002ApJ...573..199N,2010MNRAS.408..322K,
  2012A&A...541A..35D,2013MNRAS.434.2593T,2015ApJ...801..117T,2016MNRAS.457.1037W,2016A&A...592A..18M,
  2015MNRAS.452..278T,2015MNRAS.446.1175T,2015ApJ...810L..26T,2016PASA...33...10T,2017MNRAS.466.1788W,2017PASJ...69...95T,2018MNRAS.475.1859W,
  2018ApJ...868...22T,2018MNRAS.480.4434W,2019MNRAS.486.2587W,2019MNRAS.484.2119K,2020A&A...639A..86K,2020ApJ...896..158T,2020ApJ...900..180M,
  2020arXiv200907820Z,2020arXiv200907796Z}.
By contrast, when dusts grow in a cloud core or YSO,
the total surface area and number density of the dusts decrease
and the system becomes close to the ``dust-free" case.
Thus, it is expected that the specific electro-magnetic force on the dust
and the contribution of the dust to the electric current become negligible.
Simultaneously, it is expected that the electric force on the gas becomes negligible as in
equation (\ref{single_fluid_motion1}) to (\ref{single_fluid_motion2}).

Taking all these factors into account, we employ the following approximations in our numerical scheme in solving equation (\ref{single_fluid_dv2}):
\begin{eqnarray}
\label{assumption11}
|\fgp+\fgem| &=& |-\frac{\nabla P_g}{\rho_g}+\frac{1}{\rho_g}(\tau_g \eleE+ \frac{\cul_g \times \magB}{c})| \nonumber \\
&\gg& |\fdem|=|\frac{1}{\rho_d}(\tau_d \eleE+ \frac{\cul_d \times \magB}{c})|, \\
\label{assumption12}
|\fgm|= |\frac{1}{\rho_g}\frac{\cul_g \times \magB}{c}|  &\gg& |\fge|= |\frac{1}{\rho_g}\tau_g \eleE|, \\
\label{assumption13}
  |\cul_g| &\gg& |\cul_d|,
\end{eqnarray}
or equivalently,
\begin{eqnarray}
\label{assumption2}
\fgem &=& \frac{1}{\rho_g}(\tau_g \eleE+ \frac{\cul_g \times \magB}{c}) \sim \frac{1}{\rho_g}(\cul \times \magB), \\
\fdem &=& \frac{1}{\rho_d}(\tau_d \eleE+ \frac{\cul_d \times \magB}{c}) \sim 0.
\end{eqnarray}
These approximations greatly simplify the scheme and equation (\ref{single_fluid_dv2}) becomes,
\begin{eqnarray}
  \label{single_fluid_dv3}
  \frac{D \dv }{D t} &=&-\frac{\dv}{\tstop}+\frac{1}{\rho_g}[-\nabla P_g+ \frac{\cul \times \magB}{c}] \nonumber \\
  &-& (\dv \cdot \nabla)\vel+\frac{1}{2} \nabla \{(2\epsilon-1)\dv^2\}.
\end{eqnarray}

The conditions in which these
approximations are valid are investigated in the next subsection.

\subsection{Validity of the approximation on the electro-magnetic force}
\label{sec_assumption}
To obtain equation (\ref{single_fluid_dv3}) from equations (\ref{gas_single_fluid_first}) to (\ref{dust_multi_fluid_last}),
we have assumed that
\begin{enumerate}
\item the relative velocities among the charged dusts and neutral dusts
  are small and negligible (from equation (\ref{dust_multi_fluid_first}) and (\ref{dust_multi_fluid_last}) to (\ref{dust_single_fluid_first}) and (\ref{dust_single_fluid_last})).
\item the specific force acting on the gas is much larger than the electro-magnetic force acting on the dusts (equation (\ref{assumption11})).
\item the electric force of the gas phase is much smaller than the magnetic force and negligible (equation  (\ref{assumption12})).
\item the electric current of dust is much smaller than that of gas and  negligible (equation (\ref{assumption13})).
\end{enumerate}
In the rest of this section, we investigate the condition in which these assumptions are valid.
For a quick summary of the result of the validation, see \S \ref{summary_of_22}.

\subsubsection{estimate of the stopping time}
For later convenience, we estimate the stopping time here.
We assume the Epstein drag law \citep{1924PhRv...23..710E}
because mean-free-path is $\lambda_{mfp}=m_g/(\rho_g \sigma_g)\sim 195~{\rm cm} (\rho_g/(10^{-11} \gcm))^{-1}$ where
$\sigma_g=2\times 10^{-15} ~{\rm cm^2}$ and $m_g=4 \times 10^{-24} ~{\rm g}$ are assumed,
and much larger than the dust size considered in this paper.
On the assumption that the Epstein drag law is applicable,
the stopping time $t_{\rm stop}$ is given as
\begin{equation}
t_{\rm stop} = \frac{\rho_{\rm mat} a_d}{\rho_g v_{\rm therm}},
\end{equation}
where $a_d$ is the dust size, $v_{\rm therm}=\sqrt{8/\pi} c_s$ is the thermal velocity, and $\rho_{mat}$
is the internal density of the dust.
The  ratio of the stopping time to the free-fall time $t_{\rm ff}=\sqrt{ 3 \pi/(32 G \rho_g)} $ is calculated as
\begin{eqnarray}
    \label{ratio_tstop_tff}
  t_{\rm stop}/t_{\rm ff}=3.1 \times 10^{-3}
  \left(\frac{\rho_{\rm mat}}{2 \gcm}\right)
  \left(\frac{a_d}{1 \mum}\right) \nonumber\\ 
  \left( \frac{\rho_g}{10^{-18}\gcm} \right)^{-1/2}
  \left(\frac{c_{s}}{190 \ms}\right)^{-1},
\end{eqnarray}
Thus, dusts with sizes of
\begin{eqnarray}
a_d \lesssim 320 \left(\frac{\rho_g}{10^{-18} \gcm} \right)^{1/2} \mum,
\end{eqnarray}
are found to satisfy the condition of $t_{\rm stop}/t_{\rm ff} \lesssim 1$. Therefore 
the terminal velocity approximation is good for the (sub-){\rm mm} sized dust grains.

In the following subsections, we employ the terminal velocity approximation,
which allows us to investigate the conductivity, relative velocity
between the charged and neutral dusts,
electro-magnetic force of and relative velocity between the gas and dust, and electric field in detail.

\subsubsection{conductivities and electric field under terminal velocity approximation}
In order to clarify the conditions under which single-fluid approximation for the dusts and the approximations
of equations (\ref{assumption11}) to (\ref{assumption13}) 
hold, we consider a general force balance equation under
terminal velocity approximation from this subsection to \S 2.2.9.

With the terminal velocity approximation, 
the conductivities can be calculated, which is used 
to evaluate the velocity of charged species with respect to the neutrals,
electric field, and electro-magnetic force on the gas ($\fgem$) and the dust ($\fdem$)
for a given $\magB$ and $\cul$.

The velocities of the charged particles in the rest frame of the neutrals is
given as \citep[e.g.,][]{2002ApJ...573..199N,2007Ap&SS.311...35W, 2012A&A...541A..35D}
\begin{eqnarray}
\label{eq_relative_velocity}
\vel'_s&=&\frac{1}{n_s q_s} [\sigma_{O, s} \eleE'+\sigma_{H, s} {\hat \magB} \times \eleE' \nonumber \\
  &-& (\sigma_{P, s}-\sigma_{O, s}) {\hat \magB} \times {\hat \magB} \times \eleE'],
\end{eqnarray}
where the prime symbol ($'$) denotes
the physical quantity in the rest frame of the neutrals and 
\begin{eqnarray}
  \sigma_{O,s}&=&\frac{c}{B} n_sq_s\beta_s,\\
  \sigma_{H, s}&=&-\frac{c}{B} \frac{n_s q_s\beta_s^2}{1+\beta_s^2},\\
\sigma_{P, s}&=&\frac{c}{B} \frac{n_s q_s \beta_s}{1+\beta_s^2} 
\end{eqnarray}
are the Ohmic, Hall, and Pedersen conductivities, respectively,  of the charged species
(see \S \ref{chemical_reactions} for specific species considered in this paper).
The subscript $s$ denotes  the species.

By summing up the charged species in the gas phase and dust phase, respectively,
the (partial) electric current of the gas and dust are given
by 
\begin{eqnarray}
\label{generalized_ohm}
\cul'_{[g, d]}&=&\sigma_{O, [g, d]} \eleE'+\sigma_{H, [g, d]} {\hat \magB} \times \eleE' \nonumber \\
&-& (\sigma_{P, [g, d]}-\sigma_{O, [g, d]}) {\hat \magB} \times {\hat \magB} \times \eleE',
\end{eqnarray}
for both the gas and dust.
Here $\sigma_{[O, H, P], g}=\sum_{j=g} \sigma_{[O, H, P], j}$
and $\sigma_{[O, H, P], d}=\sum_{j=d} \sigma_{[O, H, P], j}$ where
$\sum_{j=g}$ and $\sum_{j=d}$ mean the summations
of the gas- and dust-phase species, respectively,
each of which is individually summed up.
As such, equation (\ref{generalized_ohm})
is a ``partial" generalized Ohm's law
for partial current of gas and dust.
Here,
\begin{eqnarray}
  \beta_s=\frac{q_s B}{m_s c \gamma_s \rho_g},
\end{eqnarray}
is the Hall parameter which is the product of the cyclotron frequency
and the collision frequency with the neutral gas.

The electric field is calculated as
\begin{eqnarray}
  \label{eleE_eq}
\eleE'=\frac{4 \pi}{c^2} (\eta_{O} \cul' + \eta_{H} \cul' \times  {\hat \magB}  +\eta_A (\cul' \times {\hat \magB}) \times {\hat \magB}).
\end{eqnarray}
where
\begin{eqnarray}
\eta_O&=&\frac{c^2}{4 \pi}\frac{1}{\sigma_O},\\
\eta_H&=&\frac{c^2}{4 \pi}\frac{\sigma_H}{(\sigma_H^2+\sigma_P^2)},\\
\eta_A&=&\frac{c^2}{4 \pi}\frac{\sigma_P}{(\sigma_H^2+\sigma_P^2)}-\eta_O,
\end{eqnarray}
are the Ohmic, Hall, and ambipolar resistivities, respectively, denoted
with subscripts of ``$O$'', ``$H$'', and ``$A$'', respectively.
The total conductivities and total current are calculated with  $\sigma_{[O, H, P]}=\sigma_{[O, H, P], g}+\sigma_{[O, H, P], d}$ and $\cul'=\cul'_g+\cul'_d$ where a pair of square brackets ([]) indicates any one of the symbols inside.

Under magnetohydrodynamics (MHD) approximation, the magnetic field,
total electric current, and electro-magnetic force are Lorentz invariant parameters.
Thus, for a given $\magB$ and $\cul$,  we can calculate $\eleE'$ using equation (\ref{eleE_eq}),
and then $\vel_s'$ using equation (\ref{eq_relative_velocity}),
$\cul_g'$ and $\cul_d'$ using equation (\ref{generalized_ohm}), and 
electro-magnetic forces on the gas ($\fgem=\fgem'=\frac{1}{\rho_g}(\tau_g' \eleE' +\frac{\cul_g' \times \magB'}{c})$)
and dust ($\fdem=\fdem'=\frac{1}{\rho_d}(\tau_d' \eleE' +\frac{\cul_d' \times \magB'}{c})$)
once the abundances of the ions, electrons and charged dusts have been determined from the chemical reaction.

In the rest of this section, we assume that the electric current is given by
\begin{eqnarray}
|\cul|=|\frac{c}{4 \pi} \nabla \times \magB| \sim \frac{c}{4 \pi} \frac{B}{\lambda_J},
\end{eqnarray}
where $\lambda_J= \sqrt{c_s^2/(2 \pi G \rho)}$ is the Jeans length and we assume
the sound speed of $c_s=190 (T/10 {\rm K})^{1/2} \ms$ with $T=10(1+\gamma (\rho/\rho_{\rm crit})^{\gamma-1}) ~K$ where $\gamma=7/5$ and $\rho_{\rm crit}= 10^{-13}~\gcm$.
We also assume that the angle between $\cul$ and $\magB$ is 45$^\circ$.

\subsubsection{chemical reactions}
\label{chemical_reactions}
To evaluate the abundances of the charged species,
we consider two types of chemistry calculations.
One is a typical chemical network calculation which is applicable for the dusts with  $\lesssim 1 \mum$. Hereafter we denote this calculation as
"the chemical network calculation" \citep[e.g.,][]{1990MNRAS.243..103U,2002ApJ...573..199N}.
The other is an analytical calculation which is applicable for the dusts with   $\gtrsim 1 \mum$.
Hereafter we denote this calculation as "the analytical calculation" \citep{2009ApJ...698.1122O}.

In the chemical network calculation, ion species of
${\rm H_2^+,H_3^+,HCO^+,Mg^+}$ 
${\rm  He^+,C^+,O^+,O_2^+,H_3O^+,OH^+,H_2O^+}$ and the neutral species of 
${\rm H,H_2, He, CO, O_2, Mg, O, C, HCO, H_2O, OH}$ are considered.
Neutral and singly,  doubly charged dusts, $g^0,~g^-,~g^+,~g^{2-},~g^{2+}$ are considered.
We take into account the cosmic-ray ionization (with a rate of $\zeta_{\rm cr}=10^{-17} s^{-1}$),
gas-phase and dust-surface recombination, and ion-neutral reactions.
The abundances of the species in chemical equilibrium are calculated.
See our previous paper  \citep{2020ApJ...896..158T} for
more detailed description about the chemical network calculations.

The chemical network calculation is not suitable to calculate ionization state of the medium with dusts of $\gtrsim 10 \mum$
because the mean dust charge $\langle Z \rangle$ of $\gtrsim 1 \mum$ is typically \citep{1987ApJ...320..803D}
\begin{eqnarray}
  \langle Z\rangle \sim - 23 (\frac{a_d}{10 \mum}) (\frac{T}{10 K})^{-1},
\end{eqnarray}
where we use the result of figure 11 of \citet{1987ApJ...320..803D} and $\lambda \equiv e^2/(k_B T)=1.7 \mum (T/(10~{\rm K}))^{-1}$.
This estimate shows that  a vast number of charged dust species should be considered in the chemical network, which is computationally demanding.

Fortunately,  however, for the medium with dusts of $\gtrsim 1 \mum$,
the dust charge distribution is well approximated by Gaussian distribution and an analytical model for ionization state is applicable.

\citet{2009ApJ...698.1122O} presents the analytical solution of the  number density of ion, electron and the dusts under the assumption
that the dust charge distribution is approximated by Gaussian distribution and ion abundance is dominated by single species.
In this section, we employ the analytical model of \citet{2009ApJ...698.1122O} to evaluate the abundances of the charged species
for the medium with dusts of $\geq 10 \mum$ in which we assume that HCO$^+$ is the dominant ion species.

In Appendix A, we compare the results using the chemical network calculation and using the analytical calculation for the case of
$a_d=1 \mum$ at which both models are applicable. The results are mutually
consistent and we conclude that the analytical calculation is suitable
to evaluate the abundances of the charged species with large dusts.

\subsubsection{relative velocity among charged and neutral dusts}
\label{rel_vel_dusts}
\begin{figure*}
  \includegraphics[clip,trim=0mm 0mm 0mm 0mm,width=50mm,,angle=-90]{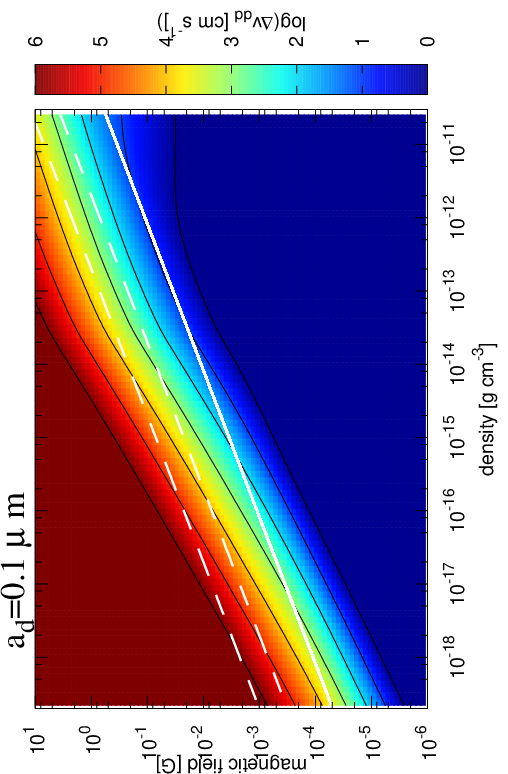}
  \includegraphics[clip,trim=0mm 0mm 0mm 0mm,width=50mm,,angle=-90]{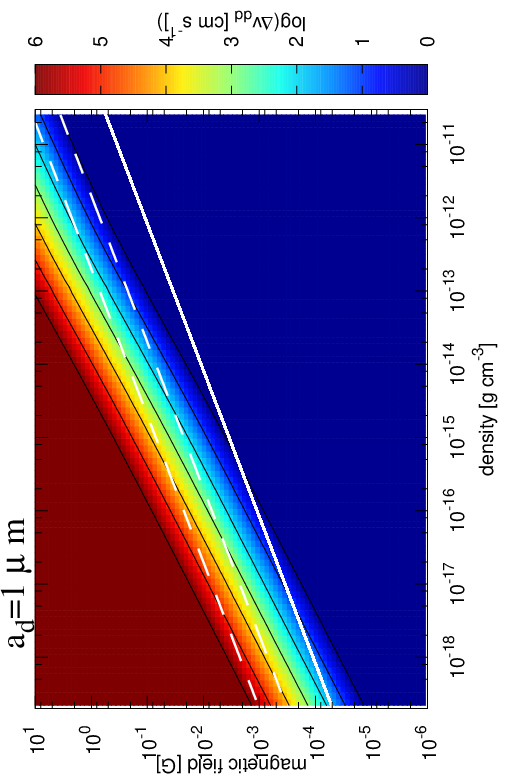}
  \includegraphics[clip,trim=0mm 0mm 0mm 0mm,width=50mm,,angle=-90]{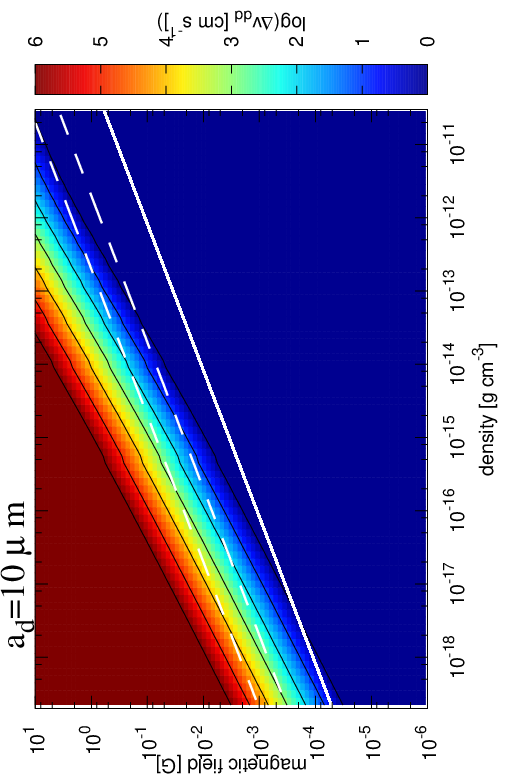}
  \includegraphics[clip,trim=0mm 0mm 0mm 0mm,width=50mm,,angle=-90]{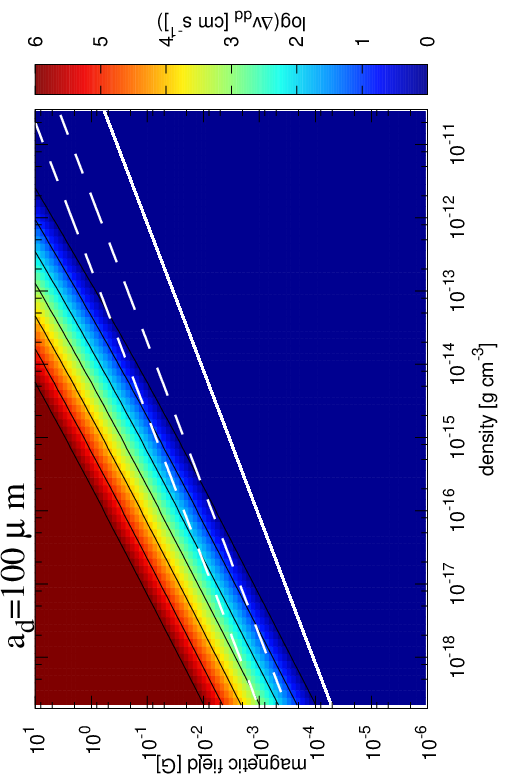}
  \caption{Relative velocity among charged dusts
    and neutral dusts, $|\dv_{\rm dd}|$ under the terminal velocity approximation on the $\rho-B$ plane.
    Top-left, top-right, bottom-left, bottom-right panels show relative velocities of the dusts of
    $a_d=0.1 \mum$, $a_d=1 \mum$, $a_d=10 \mum$, and $a_d=100 \mum$, respectively.
    For the dust size of $a_d=0.1 \mum$ and $a_d=1 \mum$, we plot the relative velocity
    between the dust with $Z=-1$ and $Z=0$.
    On the other hand, for the dust size of $a_d=10 \mum$ and $a_d=100 \mum$, we plot the relative velocity
    between the dust with $Z=\langle Z\rangle$ and $Z=\langle Z\rangle-2 \Delta Z$.
    The solid white line shows the typical magnetic field evolution
    of prestellar collapse $B=0.1  (n_g/(1~ {\rm cm^{-3}}))~\mu G$
    (main reference line) \citep{2002ApJ...573..199N}, whereas
    dashed white lines show the magnetic field
    with plasma $\beta$ of $\beta=10^{-2}$ and $\beta=10^{-3}$
    (supplementary reference lines).
    Black contours show $|\dv_{\rm dd}|=10^{0},10^{1}, \cdots, 10^{6} \cms$.
  }
  \label{relative_velocity_among_dust}
\end{figure*}

In following subsections, using the abundances of the ions and charged dusts
in chemical equilibrium obtained by the chemical network calculation and the analytical calculation,
we estimate the relative velocity among the charged dusts and neutral dusts,
the relative velocity between dust and gas, electro-magnetic forces,
and electric currents of the gas and dust
on the $\rho$-$B$ plane and show them in the figures \ref{relative_velocity_among_dust} to \ref{Jg_Jd},
and discuss the validity of the assumptions listed in \S \ref{sec_assumption}.

In the figures, we plot the critical magnetic field
\citep[thick solid line;][]{2002ApJ...573..199N} as the main reference line.
In the simulation of the collapse of the magnetized cloud core,
the gases in the envelope or circumstellar disk obey the evolution path
below the reference line \citep[e.g.,][]{2015MNRAS.452..278T} i.e.,
lower magnetic field strength for the given density,
because of the weaker initial magnetic field and magnetic diffusion.
Thus, the gases in the envelope and disk evolve in the region below the reference line.

An important exception is an outflow. In an outflow,
the magnetic field is twisted and amplified. The gas density in
it decreases as it develops \citep[see, e.g.,][]{2020ApJ...896..158T}.
As a result, the gas in the outflow could evolve toward the region above the reference line,
i.e., higher magnetic field strength for the given density.
Our previous study \citep{2020ApJ...896..158T} shows that the plasma $\beta$ in the outflow can reach
$\beta \sim 10^{-3}$. Furthermore, in practice, 3D simulation has numerical difficulties
to treat $\beta \ll 10^{-3}$ properly due to small timestep and/or numerical instability.
Thus, we consider $\beta \sim 10^{-3}$ as the practical upper limit of the magnetic field.
For these reasons, we also  plot the lines with $\beta = 10^{-3}$ and $10^{-2}$ in the figures
as the supplementary reference lines and discuss the validity of our assumption in the region
around these lines.

Figure \ref{relative_velocity_among_dust} shows the relative velocity among the charged dusts or between charged and  neutral dusts
$|\dv_{\rm dd}|$ calculated using equation (\ref{eq_relative_velocity}).
For the dust size of $a_d=0.1 \mum$ and $a_d=1 \mum$, we plot the relative velocity between the dust with $Z=-1$ and $Z=0$.
On the other hand, for the dust size of $a_d=10 \mum$ and $a_d=100 \mum$, we plot the relative velocity between the dust
with $Z=\langle Z\rangle$ and $Z=\langle Z\rangle-2 \Delta Z$ (where $\Delta Z$ is the dispersion of the charge distribution)
and hence the relative velocity between the dusts with mean charge and the dusts with
"two sigma" charge. The value of $\langle Z\rangle$ and $\Delta Z$ are $\langle Z\rangle=-23$ and $\Delta Z=2$ for $a_d=10 \mum$ and
$\langle Z\rangle=-230$ and $\Delta Z=7$ for $a_d=100 \mum$ at $T=10$ K.
Note that, for these large dusts, the abundance of neutral dusts ($Z=0$) is negligibly small.

In the case with $a_d=0.1 \mum$, the relative velocity of the charged
dusts is $|\dv_{\rm dd}| \lesssim 10^2 \cms$
in the envelope and disk  (below the main reference line) apart from
the very low density region of $\rho \lesssim 10^{-17} \gcm$, and
is only $\lesssim 1$ \% of the sound velocity. Thus, the relative velocity between
the charged and neutral dusts are very small and the charged and neutral dusts essentially move with the same velocity.
This is also the case for $a_d=1 \mum$ and $a_d=10 \mum$ in the density region $\rho >10^{-18}\gcm$.
Therefore, we can conclude that the single-fluid approximation for the dusts is
valid in the envelope and disk for the dusts of  $a_d \lesssim 10 \mum $.

However, the relative velocity can reach $|\dv_{\rm dd}| \gtrsim 10^{4} \cms$
for the dusts of  $a_d \lesssim 10 \mum $  
around $\beta\sim 10^{-3}$ (the upper supplementary reference line)
at $\rho\sim 10^{-17} \gcm$, which can be realized in the outflow region.
In this region, the velocity difference among charged dusts or between charged and neutral dusts
is not negligible and multi-fluid treatment for dusts is required.

As the dust size increases to $a_d\gtrsim 100 \mum$, the relative velocity $|\dv_{\rm dd}|$
becomes small in envelope, disk (below main reference line)
and also in outflow (around supplementary reference lines).
For example, $|\dv_{\rm dd}|\sim 10^3 \cms$ even at $\rho=10^{-18} \gcm$
  and $\beta \sim 10^{-3}$ for $a_d=100 \mum$ which is $\lesssim 5 \%$ of the sound velocity and small. 
Thus the single-fluid treatment is justified almost all situation in the YSOs
once dust size increases to $a_d \gtrsim 100 \mum$.

\subsubsection{relative velocity between the gas and the dust}
To determine the parameter range in which $|\dv_{\rm gd}|$ is sufficiently small and
  the dust is essentially completely coupled with the gas,
figure \ref{terminal_velocity} shows the terminal relative velocity between gas and dusts
$|\dv_{\rm gd}|=\tstop \{(|\fgp| + |\fgem|)-|\fdem|)\}$
of the dusts with sizes from $a_d=0.1 \mum$ (top-left panel) to $a_d=100\mum$ (bottom-right panel)
where the pressure gradient force of gas is approximated to be $\fgp=\nabla P/\rho_g \sim P/(\rho_g \lambda_J)$.
In the situation of $|\dv_{\rm gd}|$ is sufficiently small, the dust is essentially completely coupled
with the gas and the details of the electro-magnetic force on the gas and dust
is not matter for the advection of the dusts (or macroscopic dust evolution).

In the case with $a_d=0.1 \mum$, the terminal velocity of the dusts is $|\dv_{\rm gd}| \lesssim 10^3 \cms$
around the evolution path of the prestellar collapse  (below the main reference line) and
is only $\lesssim 5$ \% of the sound velocity. Thus, the dusts are completely coupled with the
gas in the envelope or disk. This is also the case for $a_d=1 \mum$
and $a_d=10 \mum$  in the density region $\rho \gtrsim10^{-18}\gcm$.
Therefore, we can conclude that  whether electro-magnetic force on the dust is negligible or not
in the envelope and disk is irrelevant in terms of for the advection of the dust of  $a_d \lesssim 10 \mum $.

However, the terminal velocity can reach $|\dv_{\rm gd}|\gtrsim 10^{4} \cms$
for the dusts of  $a_d \lesssim 10 \mum $ at 
around $\beta\sim 10^{-3}$ (the upper supplementary reference line)
at $\rho\sim 10^{-18} \gcm$.
In and around this region, the macroscopic dust drift is not negligible
and  whether electro-magnetic force on the dust is negligible or not  should be investigated.

As the dust size increases to $a_d\gtrsim 100 \mum$,  $|\dv_{\rm gd}|$
becomes large in envelope and disk (below the main reference line).
For example, $|\dv_{\rm gd}|\sim 10^4 \cms$ at $\rho=10^{-18} \gcm$ for $a_d=100 \mum$,
which is comparable to the sound velocity.
Thus the dust drift becomes nonnegligible as dust size increases to $a_d \gtrsim 100 \mum$.
The terminal velocity of such dusts also large in the outflow region
(or low-$\beta$ and low-density region which is vicinity of the supplementary reference lines).
Therefore whether electro-magnetic force on the dust is negligible or not should be investigated for
the large dusts in the almost entire region of $\rho$-$B$ plane.

\subsubsection{specific force on the gas and dust}
In this subsection, to clarify whether electro-magnetic force on the dust is negligible or not,
we investigate the ratio of the specific force on the gas and dust $|\fgp+\fgem|/|\fdem|$ .
More specifically, we investigate the condition in which the approximation of equation (\ref{assumption11}) is valid.
Note that $\fgem$ and $\fdem$ are the Lorentz invariant and satisfy $|\fgp+\fgem|/|\fdem|=|\fgp+\fgem'|/|\fdem'|$.

\begin{figure*}
  \includegraphics[clip,trim=0mm 0mm 0mm 0mm,width=50mm,,angle=-90]{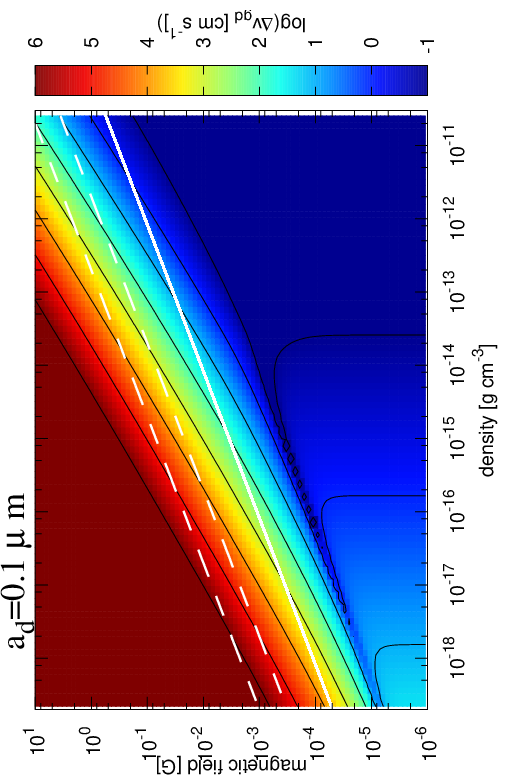}
  \includegraphics[clip,trim=0mm 0mm 0mm 0mm,width=50mm,,angle=-90]{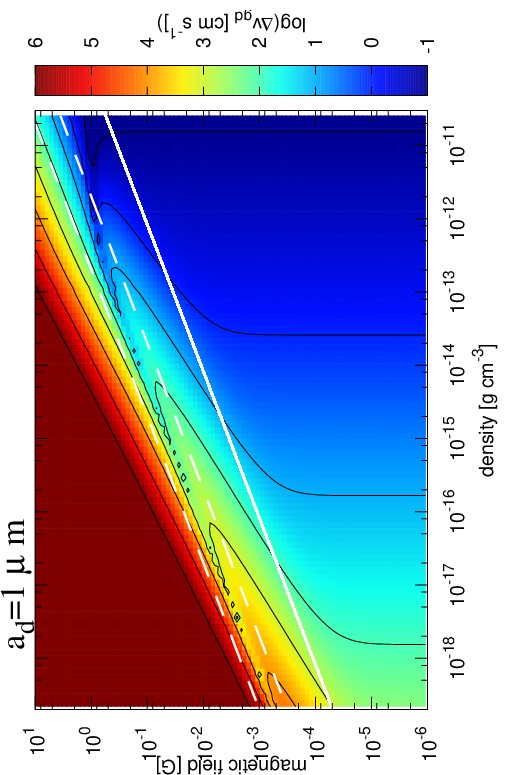}
  \includegraphics[clip,trim=0mm 0mm 0mm 0mm,width=50mm,,angle=-90]{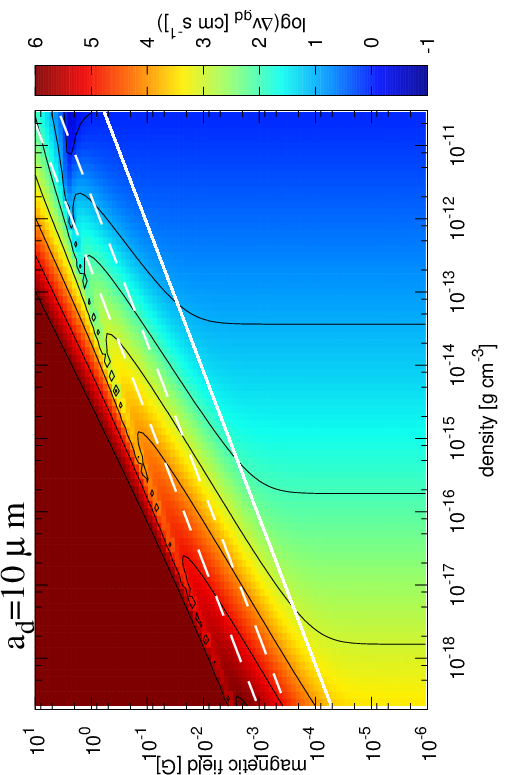}
  \includegraphics[clip,trim=0mm 0mm 0mm 0mm,width=50mm,,angle=-90]{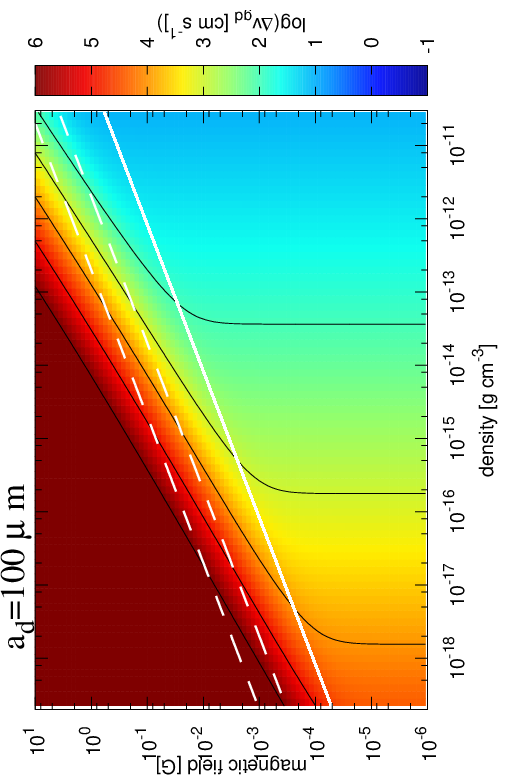}
  \caption{
    Same as figure \ref{relative_velocity_among_dust} but terminal
    relative velocity $|\dv_{\rm gd}|=\tstop [ (|\fgp| + |\fgem|)-|\fdem|]$ of the dust on the $\rho-B$ plane.
    Black contours show $|\dv_{\rm gd}|=10^{-1},10^{0}, \cdots, 10^{6} \cms$.
  }
  \label{terminal_velocity}
\end{figure*}

\begin{figure*}
  \includegraphics[clip,trim=0mm 0mm 0mm 0mm,width=50mm,,angle=-90]{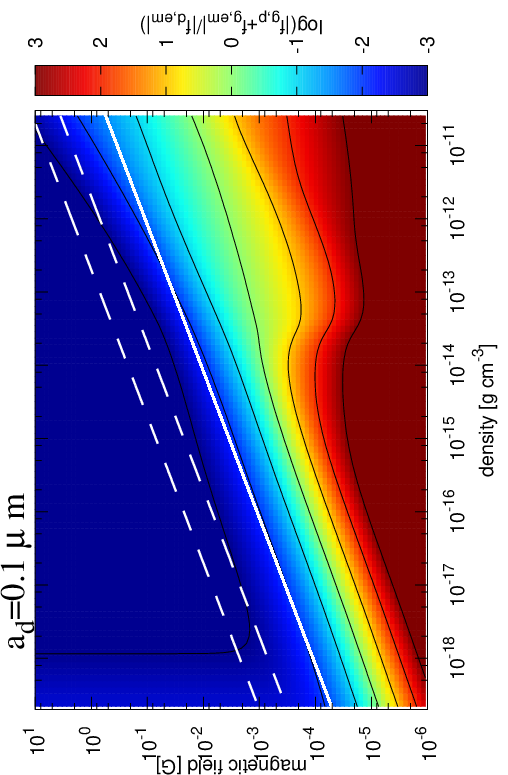}
  \includegraphics[clip,trim=0mm 0mm 0mm 0mm,width=50mm,,angle=-90]{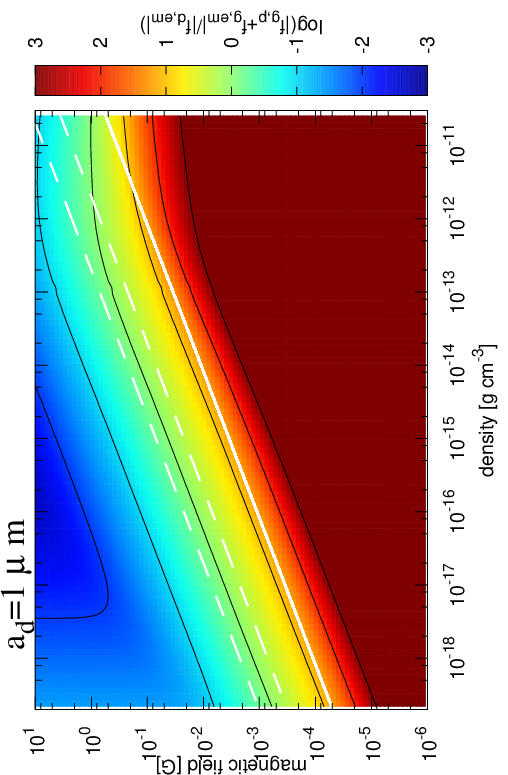}
  \includegraphics[clip,trim=0mm 0mm 0mm 0mm,width=50mm,,angle=-90]{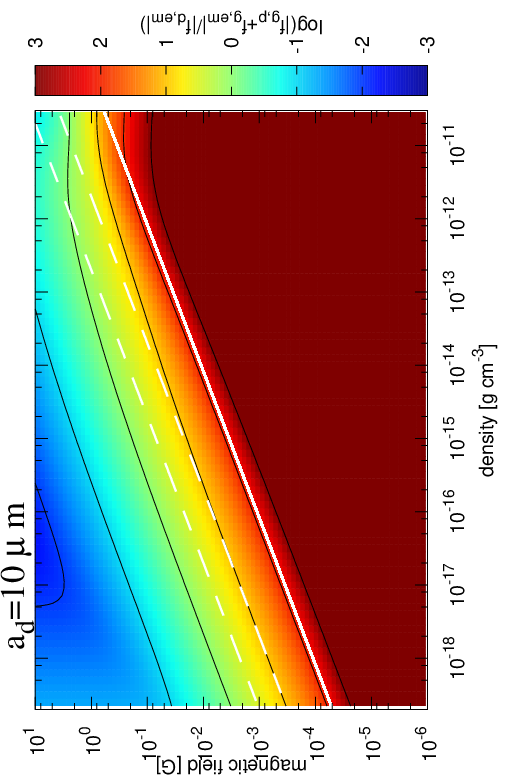}
  \includegraphics[clip,trim=0mm 0mm 0mm 0mm,width=50mm,,angle=-90]{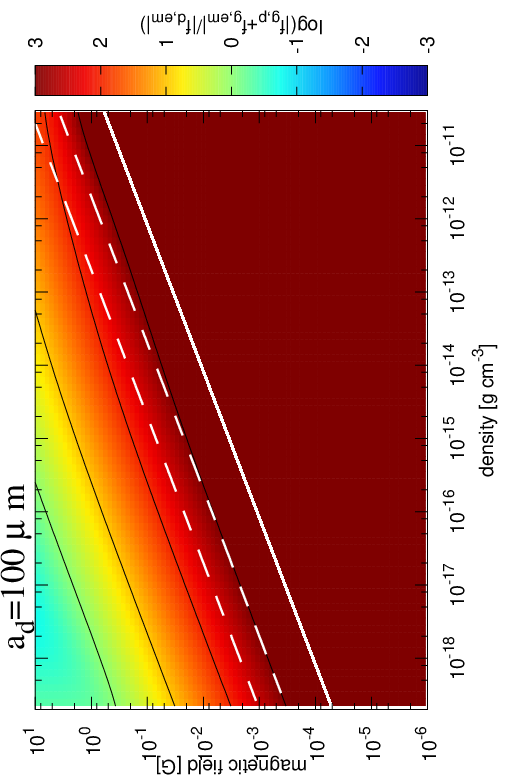}
  \caption{Same as figure \ref{relative_velocity_among_dust} but the ratio of the specific force on the gas and to
    that on the dust $|\fgp+\fgem|/|\fdem|$ is plotted on the $\rho-B$ plane.
    Black contours show $|\fgp+\fgem|/|\fdem|=10^{-3},10^{-2},\cdots,10^3$.}
  \label{fg_fd}
\end{figure*}

The top-left panel of figure \ref{fg_fd} shows $|\fgp+\fgem'|/|\fdem'|$ of $a_d=0.1 \mum$.
In the envelope and disk (below the main reference line) as
well as the outflow (supplementary reference lines), $|\fgp + \fgem|/|\fdem|<1$ for $a_d=0.1 \mum$.
This result indicates that our assumption of equation (\ref{assumption11}), meaning that
the gas receives electro-magnetic force breaks down for the dusts with $a_d=0.1 \mum$.
Note, however, that the drift velocity itself is generally
very small and negligible in the envelope and disk for $a_d=0.1 \mum$ (figure \ref{terminal_velocity}).
Therefore our assumption (the gas receive the Lorentz force and electric force is negligible)
does not affect the macroscopic dynamics (or advection)
of the dusts and hence does not result in artificial dust evolution in the envelope and disk.

However, our assumption is certainly invalid in the outflow region (around supplementary reference lines) for $a_d=0.1 \mum$.
Indeed, $|\fgp+\fgem|/|\fdem| \ll 1$ in the vicinity of the supplementary reference lines, where the terminal
drift velocity is also significant (top-left panel of figure \ref{terminal_velocity}).
Thus, our assumption does not hold in such a region.

As the dust size increases, $|\fgp+\fgem|/|\fdem|$ increases and the force on the gas gradually dominates that on the dust.
The reference line of figure \ref{fg_fd} shows $|\fgp+\fgem|/|\fdem| \sim 10^{2}$
for $a_d=1 \mum$ and  $|\fgp+\fgem|/|\fdem| \sim 10^{3}$ for  $a_d=10 \mum$  in the envelope and the disk
which implies that the assumption of $|\fgp+\fgem|/|\fdem| \gg 1$ is valid there.
However, the supplementary reference (dashed) lines show $|\fgp+\fgem|/|\fdem| \sim 1$ in the outflow region for $a_d=1,~10 \mum$.
These facts imply that our assumption of equation (\ref{assumption11})
is invalid in the outflow region for $a_d \lesssim 10 \mum$.
Furthermore, in the outflow region for $a_d \lesssim 10 \mum$,
the single-fluid approximation for the charged and neutral dusts also breaks down (figure \ref{relative_velocity_among_dust}).
Therefore, to discuss the dust dynamics of these regions, multi-fluid treatment with the
charge exchange between the gas and dusts and among the dusts is required.
Our approximation does not allow us to handle this type of dynamics
and hence it is beyond the scope of this work.

In contrast, the bottom-right panel of figure \ref{fg_fd} shows that, for the dust size of $a_d \geq 100 \mum$,
the condition of $|\fgp+\fgem|/|\fdem| \gg 1$ generally holds in the disk, envelope and the outflow.
Therefore our assumption of equation (\ref{assumption11}) is generally valid for large dusts of $a_d \gtrsim 100 \mum$.

\subsubsection{electric and magnetic force on the gas}
In the previous subsection, we have investigated the condition
in which $|\fgp+\fgem| \gg |\fdem|$ is satisfied.
In our implementation, we further assume that the magnetic force on the gas is much stronger
than the electric force on it i.e.,  $|\fgm| \gg |\fge| $ (equation (\ref{assumption12})).
Here we investigate the condition for the assumption to be valid.

Since neither $\fgm$ nor $\fge$ are Lorentz invariant, we need the following two steps.
In the first step, we investigate the condition in which 
\begin{eqnarray}
  \label{fge_fgm_ineq}
  |\fgm'|= |\frac{1}{\rho'_g}\frac{\cul'_g \times \magB'}{c}| &\gg& |\fge'|=|\frac{1}{\rho'_g} \tau'_g \eleE'|. 
\end{eqnarray}
is satisfied.
Then in the second step,  we will investigate the condition in which
\begin{eqnarray}
  \label{culg_taugv_ineq}
  |\cul'_g| &\gg& |\tau'_g \vel|.
\end{eqnarray}
is  satisfied.
When both are satisfied, our assumption is valid:
\begin{eqnarray}
  \label{fgm_ineq}
  |\fgm| \gg |\fge|.
\end{eqnarray}

To show this, let us assume that equation (\ref{culg_taugv_ineq}) holds:
\begin{eqnarray}
  |\cul'_g| \gg |\tau'_g \vel| \nonumber,
\end{eqnarray}
i.e., the bulk velocity does not contribute to the gas electric current in the rest frame of the neutral.
With this condition, we can approximate
\begin{eqnarray}
  \label{culg_taugv_approx1}
  \cul_g=\cul'_g+\tau'_g \vel \sim \cul'_g.
\end{eqnarray}
and hence, 
\begin{eqnarray}
  \cul'_g=\cul_g-\tau_g \vel \sim \cul_g, 
\end{eqnarray}
therefore,
\begin{eqnarray}
  \label{culg_taugv_approx2}
  |\cul_g| \gg |\tau_g \vel|.
\end{eqnarray}
When equation (\ref{fge_fgm_ineq}) holds,
\begin{eqnarray}
  |\frac{\cul'_g \times \magB'}{c}| &\gg& |\tau'_g \eleE'|,
\end{eqnarray}
therefore, 
\begin{eqnarray}
  \label{fgm_fge_labo1}
    |\frac{(\cul_g-\tau_g \vel) \times \magB}{c}| &\gg& |(\tau_g+ \frac{\vel \cdot \cul_g}{c^2} ) (\eleE- \frac{\vel \times \magB}{c})|  
\end{eqnarray}
where we use the Lorentz transformation of $\magB'=\magB$, $\eleE'=\eleE - \vel \times \magB/c$,
$\tau'_g=\tau_g - \vel \cdot \cul_g/c^2$, $\cul'_g=\cul_g+\tau_g \vel$ under assumptions of $v^2/c^2 \ll 1$ and
the MHD approximation being appropriate (see appendix B).

The following inequality holds from equation (\ref{fgm_fge_labo1})
for the both case of $|\tau_g| < |\frac{\vel \cdot \cul_g}{c^2}|$ and $|\tau_g| > |\frac{\vel \cdot \cul_g}{c^2}|$,
because $|\tau_g| < |\frac{\vel \cdot \cul_g}{c^2}|$ leads $|\tau_g (\eleE- \frac{\vel \times \magB}{c})| < |\frac{\vel \cdot \cul_g}{c^2} (\eleE- \frac{\vel \times \magB}{c})|$,
\begin{eqnarray}
  |\frac{(\cul_g-\tau_g \vel) \times \magB}{c}| &\gg& |\tau_g (\eleE- \frac{\vel \times \magB}{c})|,  
\end{eqnarray}
therefore,
  \begin{eqnarray}
|\frac{\cul_g \times \magB}{c} | + |\frac{2 \tau_g \vel \times \magB}{c} |  &\gg& |\tau_g \eleE  |,
\end{eqnarray}
where we use triangle and reverse triangle inequalities.
Combining equation (\ref{culg_taugv_approx2}), we finally obtain
\begin{eqnarray}
|\frac{\cul_g \times \magB}{c}|= |\rho_g \fgm| &\gg& |\tau_g \eleE| = |\rho_g \fge|.
\end{eqnarray}

\begin{figure*}
  \includegraphics[clip,trim=0mm 0mm 0mm 0mm,width=50mm,,angle=-90]{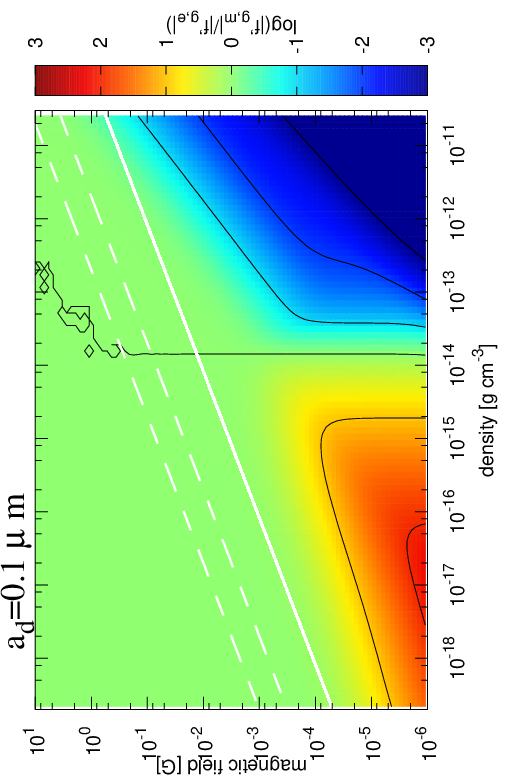}
  \includegraphics[clip,trim=0mm 0mm 0mm 0mm,width=50mm,,angle=-90]{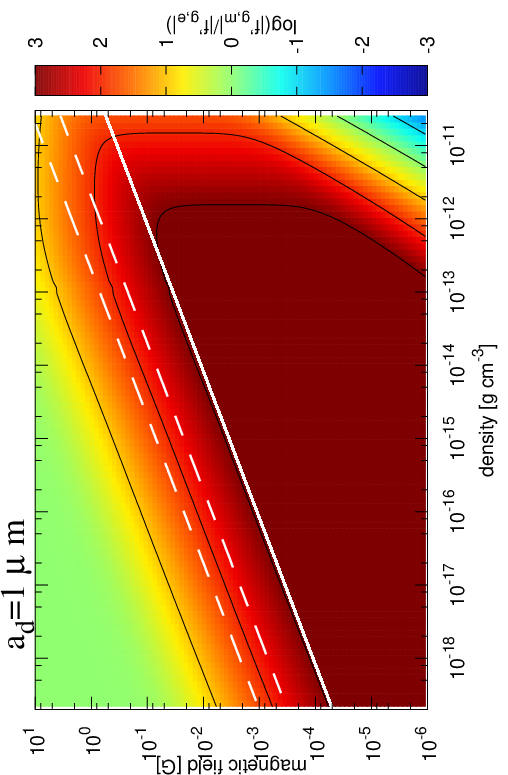}
  \includegraphics[clip,trim=0mm 0mm 0mm 0mm,width=50mm,,angle=-90]{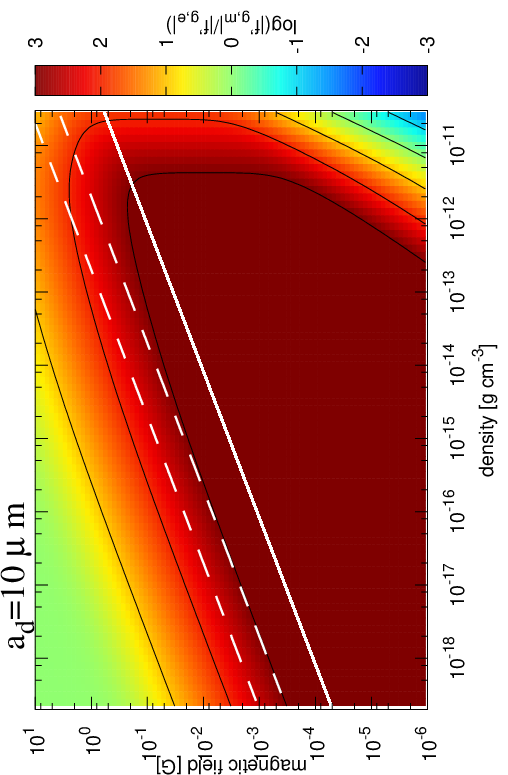}
  \includegraphics[clip,trim=0mm 0mm 0mm 0mm,width=50mm,,angle=-90]{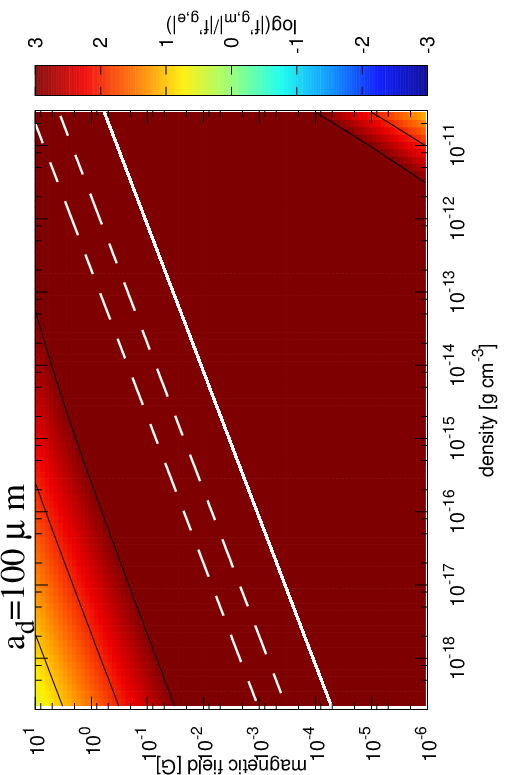}
  \caption{Same as figure \ref{relative_velocity_among_dust} but the ratio of
    the magnetic to electric forces on the gas in the neutral rest frame $|\fgm'|/|\fge'|$ is plotted on the $\rho-B$ plane.
    Black contours show $|\fgm'|/|\fge'|=10^{-3},10^{1}, \cdots, 10^{3}$.}
  \label{fgm_fge}
\end{figure*}

\begin{figure*}
  \includegraphics[clip,trim=0mm 0mm 0mm 0mm,width=50mm,,angle=-90]{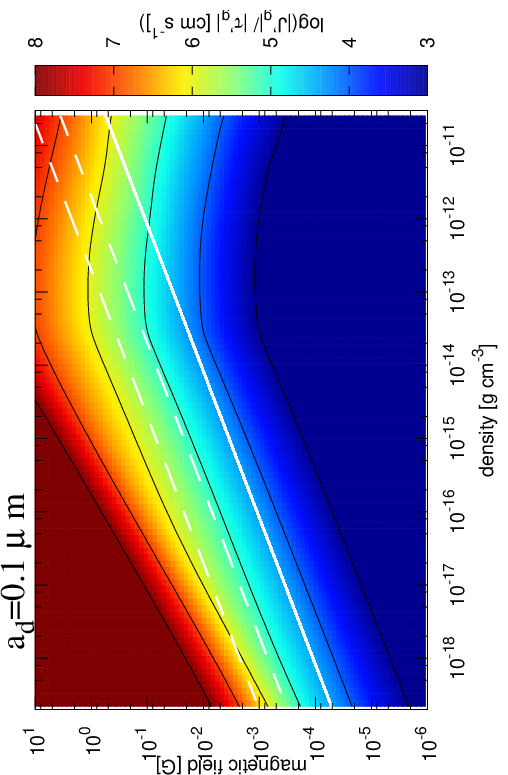}
  \includegraphics[clip,trim=0mm 0mm 0mm 0mm,width=50mm,,angle=-90]{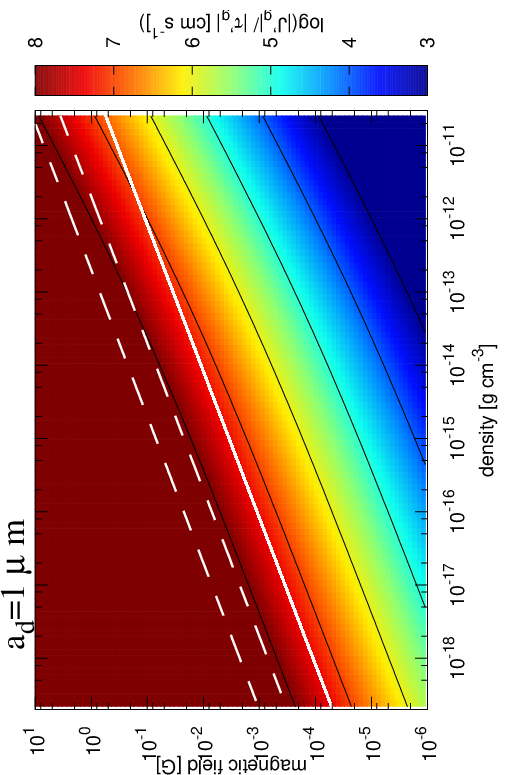}
  \includegraphics[clip,trim=0mm 0mm 0mm 0mm,width=50mm,,angle=-90]{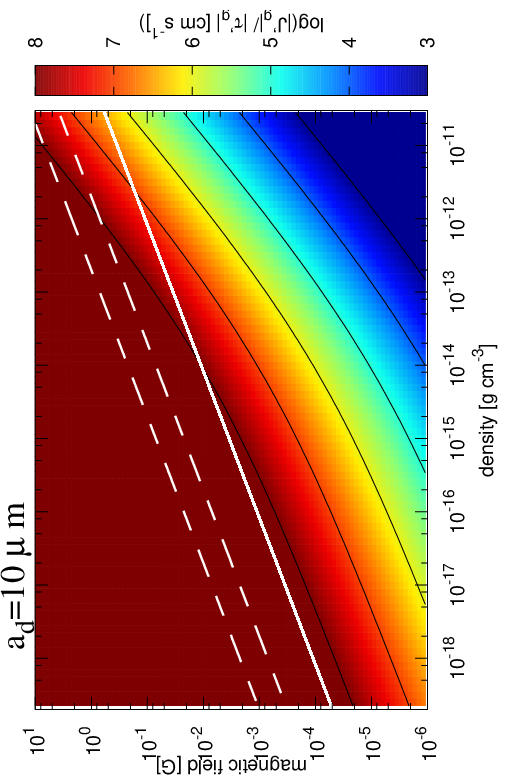}
  \includegraphics[clip,trim=0mm 0mm 0mm 0mm,width=50mm,,angle=-90]{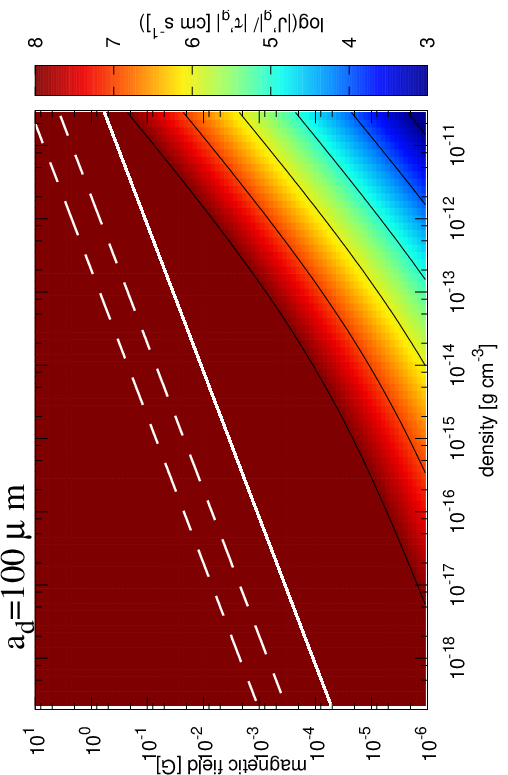}
  \caption{Same as figure \ref{relative_velocity_among_dust} but the ratio of the gas current density to the gas charge density
    $|\cul'_g|/|\tau'_g|$ is plotted on the $\rho-B$ plane.
    Black contours show $|\cul'_g|/|\tau'_g|=10^{3},10^{4},\cdots,10^8 \cms$.}
  \label{Jg_taug}
\end{figure*}

Therefore, when $|\fgm'| > |\fge'|$
(discussed below and in figure \ref{fgm_fge})
and $|\cul'_g| \gg |\tau'_g \vel|$ (discussed below and in figure \ref{Jg_taug}),
we conclude that the magnetic force of the gas phase is much larger than the electric
force of it in laboratory frame.

Figure \ref{fgm_fge} shows the ratio of the electric to magnetic forces of the
gas phase in the rest frame of the neutral  $|\fgm'|/|\fge'|$.
In the plot, for the values of $a_d=1 \mum~, 10 \mum,~ 100\mum$,
we find $|\fgm'|/|\fge'| \gtrsim 10^{2}$ in the vicinities of both
the reference lines and supplementary reference lines.
Therefore, we conclude that the magnetic force dominates
the electric force in the rest frame for $a_d\gtrsim 1 \mum$.
On the other hand, in the case of $a_d=0.1 \mum$,
the electric force is comparable to or
even larger (in high-density region) than the magnetic force and is not negligible.

Figure \ref{Jg_taug} shows $|\cul'_g/\tau'_g|$ for the same four values of $a_d$ as in
figure \ref{relative_velocity_among_dust} to \ref {fgm_fge}.
It shows that $|\cul'_g|/|\tau'_g| \gtrsim 10^{8} \cms$ for $a_d \gtrsim 100 \mum$
vicinity of the reference lines.
On the other hand, the neutral velocity $\vel$ in the disk at $r>1$ AU,  envelope, and the outflow is smaller than $10^6 \cms$.
For example the Kepler velocity is given as $1 \times 10^6 (M/0.1 \msun)^{1/2} (r/1 {\rm AU})^{-1/2} \cms$.
Thus, the relation $|\cul'_g/\tau'_g| \gg |\vel|$ holds in almost all the situations in the evolution of YSOs with $a \gtrsim 100 \mum$.
Therefore, from the result of figure \ref{fgm_fge} and \ref{Jg_taug}, we conclude $|\fgm| \gg |\fge|$ for $a\gtrsim 100 \mum$.

By contrast, for $a \lesssim 10 \mum$, the  parameter $|\cul'_g|/|\tau'_g|$ could be comparable with or smaller
than the neutral velocity.
Thus, the detailed information of electric force is required for precise computation of the dynamics for the dusts.
However, again the terminal velocity for such dusts
is very small (figure \ref{terminal_velocity}) in the disk and envelope, and
assuming the perfect coupling ($\Delta \vel_{\rm gd}=0$) is sufficient in practice in the envelope and disk.

Note that, although $|\cul'_g|/|\tau'_g|$ has a dimension of velocity, it can be regarded as the
velocity of neither the ions nor electrons because $\tau_g$ is the sum of
the positive and negative charges and is smaller than
the charge density of the ions or electrons. 
The ion and electron velocities are much smaller than $|\cul_g|/|\tau_g|$.

Note also that when the dust size is sufficiently large and when the abundance of the charged dusts sufficiently small,
the system behaves like single-fluid MHD.
In this case, one can use Maxwell's equations for $\tau_g$ and $\cul_g$ and derive
$\cul_g/(\tau_g \vel) \sim O(c \nabla \times \magB)/((\nabla \cdot \eleE) \vel ) \sim O(c \magB)/(\eleE \vel ) \sim c^2/v^2 \gg 1$.
This is consistent with the result for the dusts of  $a \gtrsim 100 \mum$.

\subsubsection{contribution of the dusts to the electric current}
Finally, we investigate the condition in which $|\cul_g| \gg |\cul_d|$ is satisfied.
Note that the condition of $|\cul_g| \gg |\cul_d|$, or equivalently $\cul \sim \cul_g$
is assumed in equation (\ref{assumption13}) and 
is also required for the dusts to deviate from the terminal velocity approximation.

The discussion in the previous subsections has relied on the terminal velocity approximation
both the ions and charged dusts.
Our scheme, however, allows the case where the dust dynamics deviates from the terminal velocity approximation.
If the contribution of the dusts to the electric current is sufficiently small,
the approximation of equation (\ref{assumption11}) and (\ref{assumption12})  remains still
valid even when the terminal velocity approximation for dust breaks down.
This is because an assumption that the {\it gas-phase} terminal velocity approximation
is appropriate is suffice for the discussion in the previous subsections to be valid.
When $\cul_g \gg \cul_d$,  the total conductivity,  total resistivity, and electric field,
are solely determined from the force balance in the gas phase and
any deviation of the dust velocity from the terminal velocity approximation does not
alter the discussion quantitatively.

\begin{figure*}
  \includegraphics[clip,trim=0mm 0mm 0mm 0mm,width=50mm,,angle=-90]{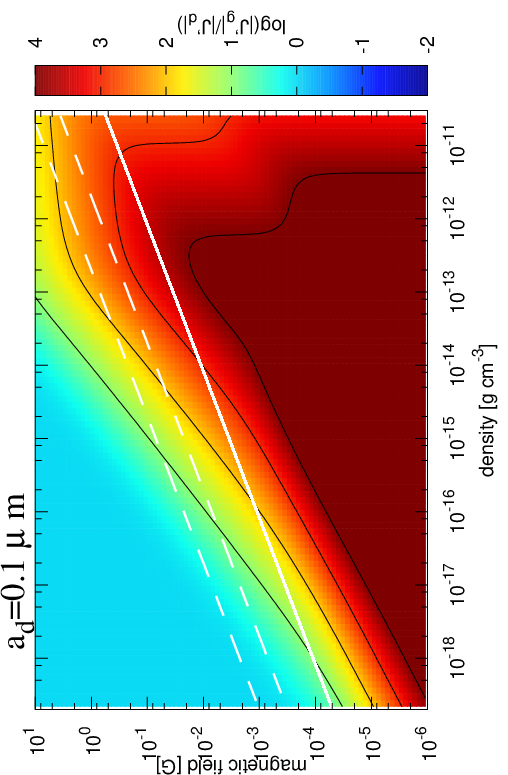}
  \includegraphics[clip,trim=0mm 0mm 0mm 0mm,width=50mm,,angle=-90]{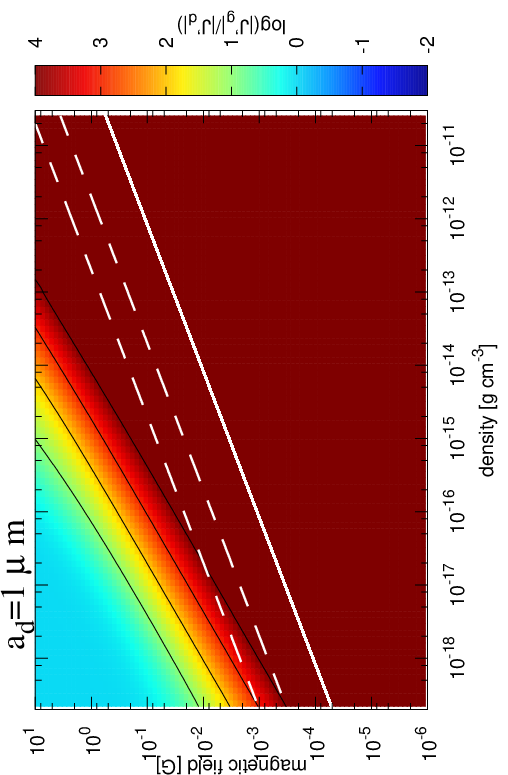}
  \includegraphics[clip,trim=0mm 0mm 0mm 0mm,width=50mm,,angle=-90]{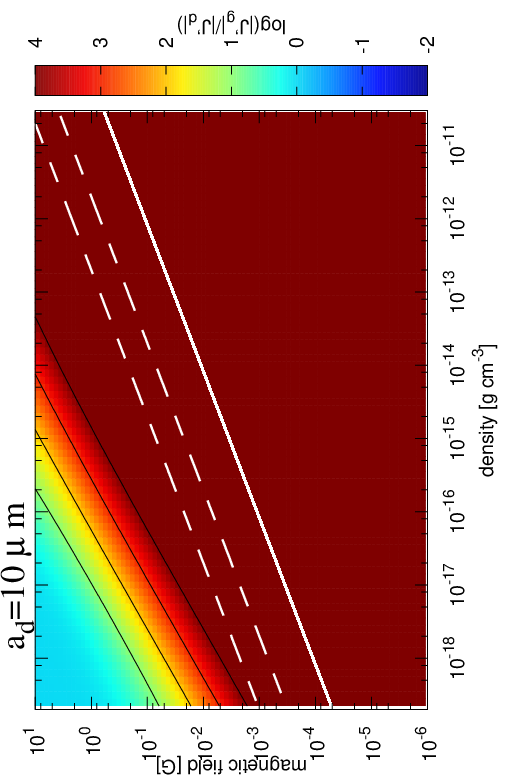}
  \includegraphics[clip,trim=0mm 0mm 0mm 0mm,width=50mm,,angle=-90]{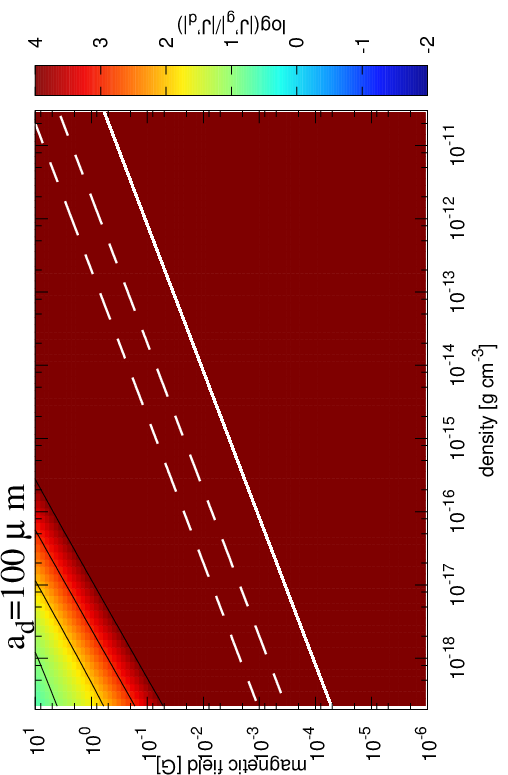}
  \caption{
    Same as figure \ref{relative_velocity_among_dust} but
    the ratio of the electric current of the gas to that
    of the dusts in the neutral rest frame $|\cul'_g|/|\cul'_d|$ is plotted
    on $\rho-B$ plane.
    Black contours show $|\cul'_g|/|\cul'_d|=10^{1},10^{2},\cdots,10^4$.
  }
  \label{Jg_Jd}
\end{figure*}

Figure \ref{Jg_Jd} shows the ratio of the  electric current of the gas to that of the  dust
$|\cul_{g}'|/|\cul_{d}'|$.
It shows that, for $a>1 \mum$, the electric current of the gas phase dominates that of dust phase,
$|\cul_{g}'|/|\cul_{d}'| \gg 1$.
For example, the result with  $a = 1 \mum$ (top-right panel)
shows $|\cul_{g}'|/|\cul_{d}'| \gtrsim 10^3$ in the vicinity of the reference lines.
Bottom panels show that the parameter $|\cul_{g}'|/|\cul_{d}'|$ increases as the dust size increases
and we find  $|\cul_{g}'|/|\cul_{d}'|>10^4$ in the almost entire region of the parameter space in the bottom panels.
Consequently, the relation $|\cul_{g}'|/|\cul_{d}'| \gg 1$ holds for $a>1 \mum$,
and hence  $|\cul_{g}|/ |\cul_{d}|\gg 1$ (it follows from $|\cul_g| \gg |\tau_g \vel|$ and
$|\cul_g| \gg |\tau_d \vel|$ {\it i.e.}, the current created by the bulk motion of the
dust $|\tau_d \vel| (=|-\tau_g \vel|) $ is smaller than $\cul_{g}$).

Note that, from equation (\ref{generalized_ohm}),
the relation $|\cul_{g}'|/|\cul_{d}'| \gg 1$ indicates that the contribution
of the dusts to the conductivities and hence resistivity is negligible 

By contrast, for $a\sim 0.1 \mum$, we find $|\cul_g'|/|\cul_d'| \sim 1$
in the low density and high magnetic field region. For example,  $|\cul_{g}'|/|\cul_{d}'| \sim 1$
at $\rho \sim 10^{-18} \gcm$ and $\beta \sim 10^{-3}$ in the top-left panel of figure \ref{Jg_Jd}.
Thus, the dusts of $a_d \sim 0.1 \mum$ have a non-negligible contribution to the conductivities and electric current.

In summary,  the approximations of $|\cul_g| \gg |\cul_d|$ are valid for $a_d \gtrsim 10 \mum$ (figure \ref{Jg_Jd})
Equation (\ref{single_fluid_dv2}) does not represent the generalized Ohm's law
unlike in the two-fluid ion-electron plasma, and the generalized Ohm's law of the gas phase with
the terminal velocity approximation can be used to consider the non-ideal MHD effect.

\subsubsection{summary of \S \ref{sec_assumption}}
\label{summary_of_22}
The results of \S \ref{sec_assumption} are summarized as follows.
\begin{enumerate}
\item For the dusts with $a \gtrsim 10 \mum$,
  \begin{enumerate}
  \item the single-fluid approximation for dusts is valid in the envelope, disk, and outflow.
  \item The electro-magnetic force of the dust phase is much weaker than
    the force of the gas phase and is negligible in the laboratory frame (figure \ref{fg_fd}).
  \item The electric force of the gas phase is much smaller than
    the magnetic force of it and is negligible in the laboratory frame (figures \ref{fgm_fge} and \ref{Jg_taug}).
  \item Hence, the approximation of equation (\ref{assumption11}) to (\ref{assumption13}) is generally valid in the envelope, disk, and outflow.
  \item The contribution of the dust to the electric current is negligible.
  \item In order to consider the non-ideal MHD effect, the generalized Ohm's law of the gas phase
    with terminal velocity approximation is applicable and dust has negligible contribution
    on resistivity.
  \end{enumerate}
\item For the dusts with $a \lesssim 10 \mum$,
  \begin{enumerate}
  \item the single-fluid approximation for dusts is valid in the envelope, disk but is not valid in the outflow.
  \item The approximation of equations (\ref{assumption11}) to  (\ref{assumption13}) is
    not always valid in the envelope, disk, or outflow (figures \ref{fg_fd}, \ref{fgm_fge}, and \ref{Jg_taug}).
  \item However, the relative velocity between the gas and the dusts
    is very small in the envelope and disk (figure \ref{terminal_velocity}) and the drift velocity of the dusts is essentially
    $\dv_{\rm gd}=0$. Hence, the approximation of equation (\ref{assumption11}) to (\ref{assumption13}) does not produce
    artificial dust drift even when the $\fgem$ and $\fdem$ are inappropriate in the envelope and disk.
  \item Our approximation does break down in the outflow in which the relative velocity between
    the neutral and charged dusts and relative velocity between the gas and charged dust are large.
    More precise treatment for the dynamics of the charged
      (and neutral) dusts is required to examine correctly the dust dynamics in the outflow.
     \item In order to consider the non-ideal MHD effect, the generalized Ohm's law
       including both the gas and dust with the terminal velocity approximation is applicable to the envelope and disk.
    \end{enumerate}
\end{enumerate}

To summarize in a sentence,
our approximation correctly describes the dust dynamics in the magnetized cloud core
apart from that in the outflow region (or low $\beta$ and low density region) with $a\lesssim 10 \mum$.
We adopt the single-fluid approximation for the dusts and
the approximation of equation (\ref{assumption11}) to (\ref{assumption13}) in the rest of this paper,
while being aware of the limitation of the approximation.

Although our chemical reaction network considered in this section is not extensive,
the potential variation due to the chemical reactions does not significantly alter the
results in this section because the conductivities and resistivities
are determined mainly by the abundance of the
major charge carriers (e.g., HCO$^+$) and therefore adding minor species into the
chemical network does not significantly affect them \citep[see, e.g.,][]{2018MNRAS.478.2723Z}.
Note also that we only investigate the case of  $a_d \geq 0.1 \mum$ and we should not extrapolate our results to
  very small dusts such as $a_d \sim 0.005 \mum$ which is lower limit of MRN model.

Our conclusion that the Lorentz force on the dust is negligible for $a_d \gtrsim 10 \mum$ in the envelope and disk is 
consistent with the conclusion of \citet{2017MNRAS.469.3532L}.
\citet{2017MNRAS.469.3532L} study the dynamics of charged dusts
in giant molecular clouds (GMCs) and shows that dusts with size of $a_d>1 \mum$ essentially behave
as neutral particles. Their result strengthen our conclusions because, even
in a dilute gas like the GMC, the Lorentz force is negligible for $a_d>1 \mum$.

\subsection{Generalized Ohm's law, induction equation, and energy equation}
In the rest of this paper, the equations (\ref{assumption11}) to (\ref{assumption13}) are assumed.

In our scheme, the generalized Ohm's law of 
\begin{eqnarray}
  \eleE&=&-\frac{\vel \times \magB}{c} + \eta_{O} \cul + \eta_{H} \cul \times  {\hat \magB}  +\eta_A (\cul \times {\hat \magB}) \times {\hat \magB} \nonumber \\
  &=&-\frac{\vel \times \magB}{c} + \eleE'
\end{eqnarray}
with the terminal velocity approximation is used.
Here, we assume that the contribution of the dusts
to the resistivity and current is negligible or
that the terminal velocity approximation is valid also for the dusts.
Since one of these conditions is always true (see the previous subsection),
the resistivities can be calculated in the ordinary way.
The barycentric velocity is approximated to be the neutral gas velocity.

The induction equation is then given by
\begin{eqnarray}
  \frac{D \magB}{Dt}&=&-\magB (\nabla \cdot \vel)+(\magB\cdot \nabla)\vel \nonumber \\
  &+& c \nabla \times \{  \eta_{O} \cul + \eta_{H} \cul \times  {\hat \magB}  +\eta_A (\cul \times {\hat \magB}) \times {\hat \magB}  \}
\end{eqnarray}

In our scheme, the specific total energy is numerically integrated.
The energy equation is given by
\begin{eqnarray}
  \label{equation_total_energy}
  \frac{D e_g}{Dt}&=&\frac{1}{(1-\epsilon)\rho} \nabla \cdot [\{-(P_g + \frac{\magB^2}{8 \pi}) \mathbb{I} - \frac{\magB \magB}{4 \pi}\} \cdot (\vel-\epsilon \dv) \nonumber \\
    &+& \frac{c}{4 \pi} \eleE' \times \magB ] +  \frac{\cul \cdot \eleE'}{(1-\epsilon)\rho c} - \epsilon (\Delta v \cdot \nabla) e +\epsilon \frac{\Delta \vel^2}{\tstop}.
\end{eqnarray}
where $e_g \equiv \frac{1}{2} v_g^2+u_g+\frac{B^2}{8 \pi \rho_g}$ is the specific total energy of the gas and
$u_g$ is the specific internal energy  which is calculated as $u_g=e_g- \frac{1}{2} v_g^2+\frac{B^2}{8 \pi \rho_g}$ in our scheme.
Equation \ref{equation_total_energy} is the total energy equation of the gas, and we just rewrote the Lagrangian derivative of $\vel_g$,
{\it i.e.,} $\frac{D_g}{Dt}$ to that of barycentric velocity $\vel$, {\it i.e.,}  $\frac{D}{Dt}$.
Note that we have implicitly used the equations
(\ref{assumption11}) to (\ref{assumption13}), assuming that the gas receives all the magnetic force.

\subsection{Order estimate of the $\dv \cdot \nabla \vel$ term and simplification}
\label{neglect_non_linear_term}
We estimate the orders of ratios of  $\dv \cdot \nabla \vel$
and $\dv^2$ to $\fgp+\fgem$ to be, by further simplifying equation (\ref{single_fluid_dv3}),
\begin{eqnarray}
  \frac{\dv \cdot \nabla \vel}{(\fgp+\fgem)}=O(\frac{|\dv| |\vel|/L}{(c_s^2+v_{A}^2)/ L })=O(\frac{|\dv| |\vel |}{(c_s^2+v_{A}^2)}),\\
  \frac{\nabla \dv^2}{(\fgp+\fgem)}=O(\frac{\dv^2/L}{(c_s^2+v_{A}^2)/ L })=O(\frac{\dv^2}{(c_s^2+v_{A}^2)}).
\end{eqnarray}
where $v_{A}$ is Alfv\'en velocity.
Therefore, in the sub-sonic or sub-alfvenic dust motion, which is our primary interest,
$\dv \cdot \nabla \vel$ and $\nabla \{ (2 \epsilon -1 ) \dv^2 \}$ are much smaller than $\fgp+\fgem$.
Consequently, we ignore these terms in equation (\ref{single_fluid_dv3}) in our scheme.


\section{Implementation}
In this section, we describe our implementation of the scheme.
\subsection{Discretized equations}
The smoothed particle hydrodynamics (SPH)
discretization of the  equations with regard to the dust dynamics is given by
\begin{eqnarray}
  \label{single_fluid_desc_eps}
  \frac{D \epsilon_i}{D t} &=& -\sum_j m_j [\epsilon_i (1-\epsilon_i)\frac{\dv_i \cdot \nabla W_{ij}(h_i)}{\Omega_i \rho_i} \nonumber \\
    &+&\epsilon_j (1-\epsilon_j)\frac{\dv_j \cdot \nabla W_{ij}(h_j)}{\Omega_j \rho_j} \nonumber \\
    &+& \frac{\alpha_\epsilon v_{\rm sig,\epsilon}}{\bar{\rho}}(\epsilon_i-\epsilon_j) \mathbf{e}_{ij}\cdot \bar{\nabla W_{ij}} ],\\
    \label{single_fluid_desc_v}
  \frac{D \vel_i }{D t} &=&  \frac{D \vel }{D t}|_{GSPMHD} -\sum_j [\epsilon_i (1-\epsilon_i)\dv_i \frac{\dv_i \cdot \nabla W_{ij}(h_i)}{\Omega_i \rho_i} \nonumber \\
    &+&\epsilon_j (1-\epsilon_j)\dv_j \frac{\dv_j \cdot \nabla W_{ij}(h_j)}{\Omega_j \rho_j}], \\
  \label{single_fluid_desc_dv}
  \frac{D \dv_i }{D t} &=& -\frac{\dv_i}{\tstop} -\frac{1}{1-\epsilon_i} \frac{D \vel }{D t}|_{GSPMHD}, \\
  \label{single_fluid_desc_energy}
  \frac{D e_i}{Dt}&=& \epsilon_i \frac{\dv_i^2}{\tstop}+ \frac{1}{1-\epsilon_i}\frac{D e}{Dt}|_{GSPMHD} \nonumber \\
  &-& \sum_j m_j [\epsilon_i (e_i-e_j) \frac{\dv_i \cdot  \nabla W_{ij}(h_i)}{\Omega_i \rho_i}],
\end{eqnarray}
where the $\frac{D}{Dt}_{GSPMHD}$ terms are the discretization of the Godunov SPMHD scheme \citep{2011MNRAS.418.1668I}
using the gas velocity.
Our discretization procedure for $D\epsilon_i/Dt$ and  $D\dv_i/Dt$ is almost identical to that of \citet{2014MNRAS.440.2147L}
with some modification.
In our implementation, we ignore the non-linear terms of equation (\ref{single_fluid_dv3})
(see \S \ref{neglect_non_linear_term}), which improves the solution of the dusty MHD shock-tube problem
(\S \ref{dusty_MHD_shocktube_problem}) in the weak coupling limit.
We add also the numerical dissipation term for $D\epsilon_i/Dt$
(the last term of equation (\ref{single_fluid_desc_eps})) where $\alpha_\epsilon=1$ and $v_{\rm sig, \epsilon}=1/2(c_{s,i}+c_{s,j})$.
Our numerical tests showed that the choice of $\alpha_\epsilon$ and $v_{\rm sig, \epsilon}$ is fairly conservative
and more aggressive choice for these values is possible.
In principle, numerical dissipation is required for any conserved variable
which satisfies $\sum_j m_j DA_j/Dt=0$ as pointed out by \citet{2008JCoPh.22710040P}.
In this case, $\epsilon$ is a conservative variable, as pointed out in section 3.3 of \citet{2014MNRAS.440.2147L}.
Our numerical experiments show that the dissipation term greatly stabilizes the system at discontinuity of $\epsilon$.
$D\epsilon_i/Dt$ is integrated with the second-order leap-flog method while
$D \dv_i /D t$ and $D e_i /D t$ are calculated as described in the following subsection
because of they include the friction terms.

\subsection{Implementation of friction terms}
To implement friction term, we adopt so-called piece-wise exact solution
method \citep{2008ApJ...687..303I} to remove the restriction on the timestep
in  strong coupling limit ($\Delta t/\tstop \to \infty$).

The analytical solution of equations (\ref{single_fluid_desc_dv}) and (\ref{single_fluid_desc_energy})
with regard to $\dv_i$ and $e_i$ at $t+\dt$ on an assumption that 
the terms other than the friction terms are constant is \citep{2014MNRAS.440.2147L},
\begin{eqnarray}
  \label{exact_dv}
  \Delta \vel^{n+1}&=&\dv^n \exp(-\dt/\tstop) \nonumber \\
  &-& \acc_i \tstop(1-\exp(-\dt/\tstop)),  \\
  \label{exact_energy}
  e^{n+1}&=&e^n +\frac{1}{2} \epsilon \{2 \acc_i^2 \dt - \exp(-2 \dt/\tstop) \nonumber \\
  (1&-&\exp(-\dt/\tstop)(\dv^n - \acc_i \tstop) \nonumber\\
  (\dv^n&-&\acc_i \tstop +\exp(-\dt/\tstop)(\dv^n+3 \acc_i \tstop)))\}, \nonumber\\
\end{eqnarray}
where $f^n$ and $f^{n+1}$ denote the values at $t$ and $t+\dt$, respectively and
$\acc_i=\frac{1}{2}( (\frac{D \dv}{Dt})^n + \left(\frac{D \Delta \vel}{Dt} \right)^{n+1}) $
is the force terms excluding the friction term in equations (\ref{single_fluid_desc_dv}) and (\ref{single_fluid_desc_energy}).

We find that the computation of equations (\ref{exact_dv}) and (\ref{exact_energy})
is considerably affected by the numerical artifact due to the round-off error
in the weak coupling limit ($\Delta t/\tstop \to 0$).
To avoid the numerical artifact, we rewrite these equations as
\begin{eqnarray}
  \Delta \vel^{n+1} &=& \dv^n \exp(-\dt/\tstop) - \acc_i \tstop \expm(-\dt/\tstop), \\
  e^{n+1} &=& e^n + \epsilon \{-\frac{1}{2} \dv^2 \expm(-2 \dt/\tstop)+\acc_i^2  \tstop \dt \nonumber\\
  &+& (\acc_i \cdot \dv) \tstop \expm(\dt/\tstop)^2 \nonumber\\
  &-& \frac{1}{2} \acc_i^2 \tstop^2 \expm(\dt/\tstop)(\expm(\dt/\tstop)-2), \nonumber\\
\end{eqnarray}
and employ them in our simulation code, where
the $\expm(x) (=(\exp(x)-1)) $ function in numerical library is used.
This devising allows us to calculate the correct results
up to $\Delta t/\tstop \gtrsim 10^{-16}$, the threshold of which is 
the round-off error of the double precision.
Furthermore, we use the following asymptotic expressions
for $\Delta t/\tstop < 10^{-12}$
to mitigate the round-off error for $\Delta t/\tstop \lesssim 10^{-16}$:
\begin{eqnarray}
  \dv^{n+1}&=&\dv^n - \dv^n \frac{\dt}{\tstop}+\acc_i \dt,  \\
  e^{n+1}  &=& e^n + \epsilon \dv^2 \frac{\dt}{\tstop}.
\end{eqnarray}
Note that this pair of formulae have been already derived by \citet{2014MNRAS.440.2147L}
(equation (83) and (87) \footnote{There was a typographic error in their equation (87)}).
With these numerical treatment, we can correctly calculate dust evolution in the  weak coupling limit.

\section{Results}
In this section, we present the results of test problems.
The first test is a hydrodynamical test to confirm that
our implementation is consistent with previous studies.
The second and third tests are magnetohydrodynamical tests.

\subsection{dusty wave problem}
The first test problem is the dusty wave problem, which was presented in \citet{2011MNRAS.418.1491L}.
By linearizing equations (\ref{gas_single_fluid_first}), (\ref{gas_single_fluid_last}),  (\ref{dust_single_fluid_first}), and (\ref{dust_single_fluid_last}) on assumptions of 
$\rho_g=\rho_{g,0}+ D_g \exp(i (kx-\omega t))$,
$\rho_d=\rho_{d,0}+ D_d \exp(i (kx-\omega t))$,
$v_g=V_g \exp(i (kx-\omega t))$, and
$v_d=V_d \exp(i (kx-\omega t))$,
we obtain
\begin{eqnarray}
\begin{pmatrix}
-i \omega + \frac{K}{\rho_{g,0}}&  -\frac{K}{\rho_{g,0}} & \frac{i c_s^2 k}{\rho_{g,0}} & 0 \\
-\frac{K}{\rho_{d,0}}&  -i \omega -\frac{K}{\rho_{d,0}} & 0 & 0 \\
i k \rho_{g,0} & 0 & -i \omega & 0 \\
 0 & i k \rho_{g,0} & 0 & -i \omega \\
\end{pmatrix}
\begin{pmatrix}
V_g  \\
V_d  \\
D_g  \\
D_d \\
\end{pmatrix}
=0
\end{eqnarray}
where $K=\rho_g \rho_d/\{\tstop(\rho_g+\rho_d)\}$ is the drag coefficient.
The corresponding dispersion relation is
\begin{eqnarray}
  \label{disp_relation}
  &\frac{\omega}{\rho_{g,0} \rho_{d,0}}&[c_s^2 k^2 \rho_{g,0} (-i K - \rho_{d,0} \omega) \nonumber \\
    &+&\omega^2 (i K (\rho_{g,0}+\rho_{d,0})+ \rho_{g,0}\rho_{d,0} \omega)] =0.
\end{eqnarray}
This contains two damping oscillation modes and one pure damping mode.
Here we use a damping oscillation mode propagating to the right direction as a test problem.
Since the analytical form of $\omega$ are complicated,  we calculated it numerically for given $K$.
Table 1 tabulates the calculated values
of $\omega$ and $(D_g,D_d,V_g,V_d)$ for $K=10^{-2},10^{-1},1 ,10^{1},$ and $10^{2}$
of the damping oscillation mode, where the wave number and sound velocity are fixed at $k=2 \pi$ and $c_s=1$, respectively.

We find that the eigenfrequency asymptotically obeys
\begin{eqnarray}
\omega&=&2 \pi-K/(2 \rho_d) i ~(K \to 0), \\
\omega&=&  \sqrt{2} \pi   ~(K \to  \infty).
\end{eqnarray}

\begin{figure*}
  \includegraphics[clip,trim=0mm 0mm 0mm 0mm,width=220mm,,angle=-90]{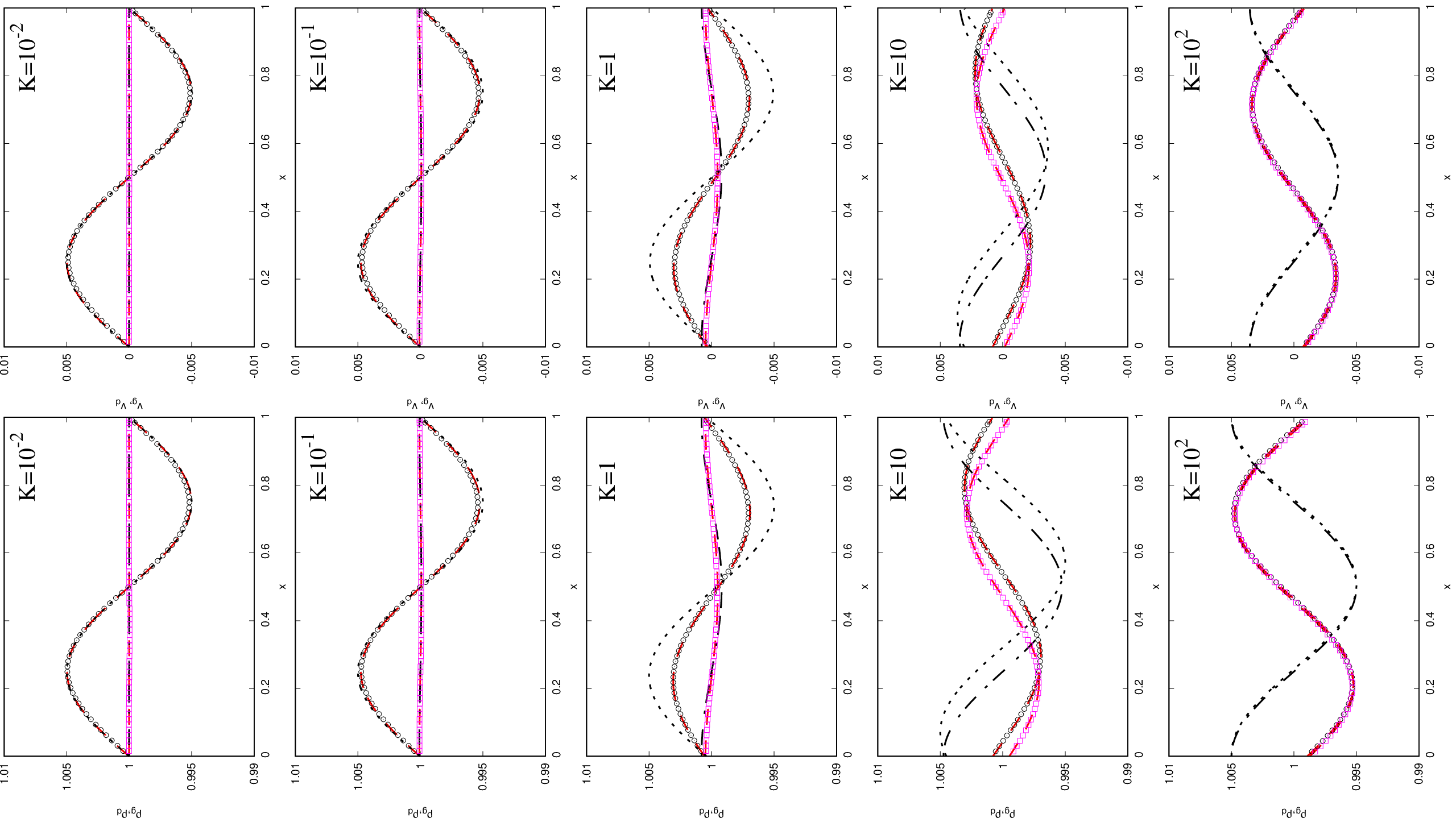}
\caption{
  Results of the dusty wave problem with five values of $K$ from $K=10^{-2}$ to $K=10^{2}$ at $t=1$.
  Black circles show the gas density (left panels) and gas velocity (right panels).
  Magenta squares show the dust density (left panels) and dust velocity (right panels).
  Red dashed lines show the analytical solutions at $t=1$.
  Black dotted and dashed-dotted lines show the initial conditions of the gas and dust, respectively.
  64 particles among the 512 particles simulated are plotted for ease of viewing.
}
\label{dusty_wave}
\end{figure*}

Figure \ref{dusty_wave} shows the results of the dusty wave problem for various values of $K$, 
where the number of the particles is 512 in the simulation domain,
the unperturbed state is given as $\rho_g=\rho_d=1$ and  $v_g=v_d=0$, and
the perturbation is a sinusoidal wave with wave length of 1 with the amplitude for
density, 0.5 \%.
The figure shows that the result of our numerical simulation agrees with the analytical solutions for any $K$,
from weak coupling limit $K \ll 1 $ to strong coupling limit $K \gg 1$,
thus confirming that our implementation correctly
reproduces the propagation of the dusty wave regardless of the coupling state and hence is valid.

The solution of the gas components with  $K \ll 1$  corresponds to the sound wave without dust friction.
By contrast, in $K \gg 1$, dust-gas mixture behaves as a single-fluid and
the solution corresponds to the sound wave of the dust-gas mixture.
Their phase velocity is slower than $c_s=1$ because of its large density (see table 1).

\begin{table*}
\begin{center}
  \caption{
    Eigenfrequency $\omega$ and eigenvectors of a damping-oscillation mode of equation (\ref{disp_relation}) for a given $K$.
}		
\begin{tabular}{cccccc}
\hline\hline
 $K$  & $\omega$ &  $V_g/D_d$ &  $V_d/D_d$ &  $D_g/D_d$ &  $D_d/D_d$   \\
\hline
$10^{-2}$   & $6.283 - 4.999 \times 10^{-3} i$ &  $ 2.533 \times 10^{-6} - 6.283 \times 10^{2} i$ & $ 0.9999 - 7.957 \times 10^{-4} i$  & $ 0.5000 - 6.283 \times 10^{2} i$  &  $1$   \\
$10^{-1}$   & $6.282 - 4.999 \times 10^{-2} i$  & $-2.533 \times 10^{-4} + 62.82 i$  & $-0.9998 + 7.955 \times 10^{-3} i$ & $-0.5001 + 62.82 i$ &  $1$   \\
$1$        & $6.184 - 0.4870 i$     & $2.556 \times 10^{-2} - 6.126 i$  & $0.9842 - 7.751 \times 10^{-2} i$  &  $0.5130 - 6.184 i$ &  $1$   \\
$10^{1}$    & $4.530 - 0.4921 i$  & $0.6500 - 0.4010 i$  & $0.7209 - 7.833 \times 10^{-2} i$  & $0.9508 - 0.4530 i$  &  $1$   \\
$10^{2} $   & $4.444 - 4.935 \times 10^{-2} i$  &  $0.7065 - 3.928 \times 10^{-2} i$  & $0.7073 - 7.854 \times 10^{-3} i$  &  $0.9995 - 4.444 \times 10^{-2} i$  &  $1$  \\
\hline
\end{tabular}
\end{center}
\footnotesize
\label{eigen_frequency}
\end{table*}

\subsection{dusty MHD shocktube problem}
\label{dusty_MHD_shocktube_problem}
Next, we investigate a one-dimensional dusty MHD shocktube problem.
Since the analytical solution of the dusty MHD shocktube is not known,
we use an initial condition in which the dust back-reaction is negligible
and compare the numerical results with a known solution of the MHD shocktube.

Our initial conditions are given
as $(\rho_g,P,v_{g, x},v_{g, y},v_{g, z},B_y,B_z)=(1.08,0.95,1.2,0.01,0.5,3.6/\sqrt{4 \pi},2/\sqrt{4 \pi})$ for $x<0$ and
$(1,1,0,0,0,4/\sqrt{4 \pi},2/\sqrt{4 \pi})$ for $x>0$, the same as those
presented in \citet{1994JCoPh.111..354D} and \citet{1995ApJ...442..228R}.
The exact solution consists of two fast shocks, two rotational discontinuities, two slow shocks,
and one contact discontinuity.
The initial values of $\epsilon$ and $\vel_d$ are $\epsilon=10^{-3}$ and  $\vel_d=0$, respectively.
Note that a small value is adopted for $\epsilon$ to make the dust back-reaction negligible
and exact solution for the gas applicable.
The number of the particles in the simulation domain is 830.

We investigate two extreme cases,  i.e., the weak and the strong coupling limit
both of which have obvious exact solutions.
In the strong coupling limit,
the exact solution is $\vel_d=\vel_g$ and $\rho_d=\epsilon \rho$;
this is a relatively straightforward problem
in our scheme because $\dv$ converges to  $\dv=0$ due to the friction.
In the  weak coupling limit, on the other hand,
$\vel_d$ and $\rho_d$ keep their initial profiles; this
is a difficult problem in our scheme because
$\dv$ should exactly cancel the gas velocity ($\dv=-\vel_g$).

\begin{figure*}
  \includegraphics[clip,trim=0mm 0mm 0mm 0mm,width=100mm,,angle=-90]{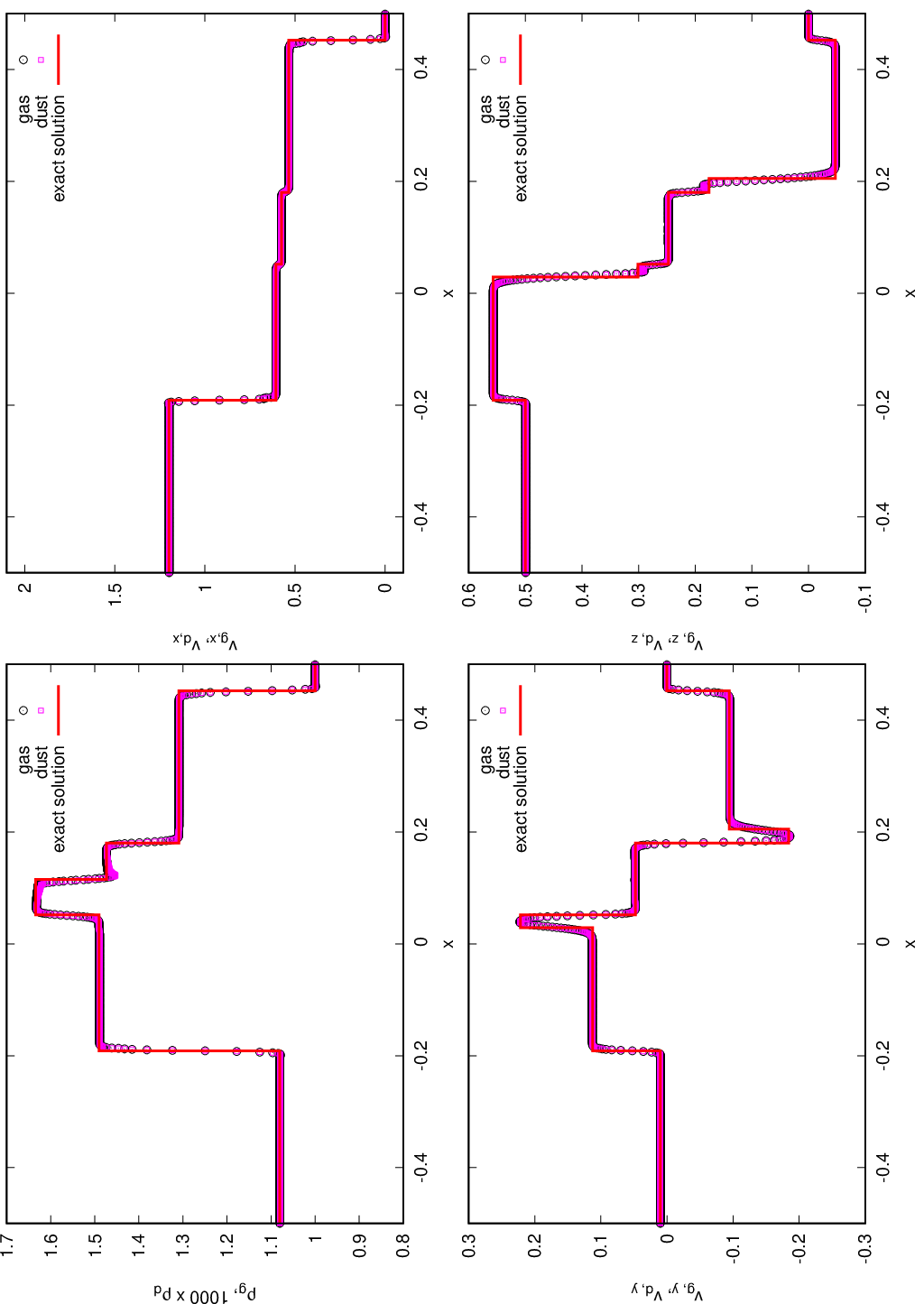}
\caption{
  Results of the dusty MHD shock tube problem in the strong coupling extreme ($K=10^4$).
  Black circles and magenta squares show the gas and dust respectively.
  Top-left panel shows $\rho_g$ and  $\rho_d/\epsilon=1000 \rho_d$.
  Top-right panel shows $v_{g, x}$ and  $v_{d, x}$.
  Bottom left panel shows  $v_{d, y}$, and $v_{g, y}$.
  Bottom right panel shows $v_{g, y}$ and  $v_{d, y}$.
  Red lines show the analytical solutions for the gas and dust.
}
\label{dusty_MHD_shock_strong}
\end{figure*}

Figure \ref{dusty_MHD_shock_strong}
shows the results (density and velocity)
of the dusty MHD shocktube problem in strong coupling extreme ($K=10^4$).
The profile for the dust and gas are indistinguishable for both the density and
velocity. 
The results are found to agree with the exact solution of the MHD shock tube problem (red lines in the figure);
it is expected because the dust inertia is sufficiently small.

\begin{figure*}
  \includegraphics[clip,trim=0mm 0mm 0mm 0mm,width=100mm,,angle=-90]{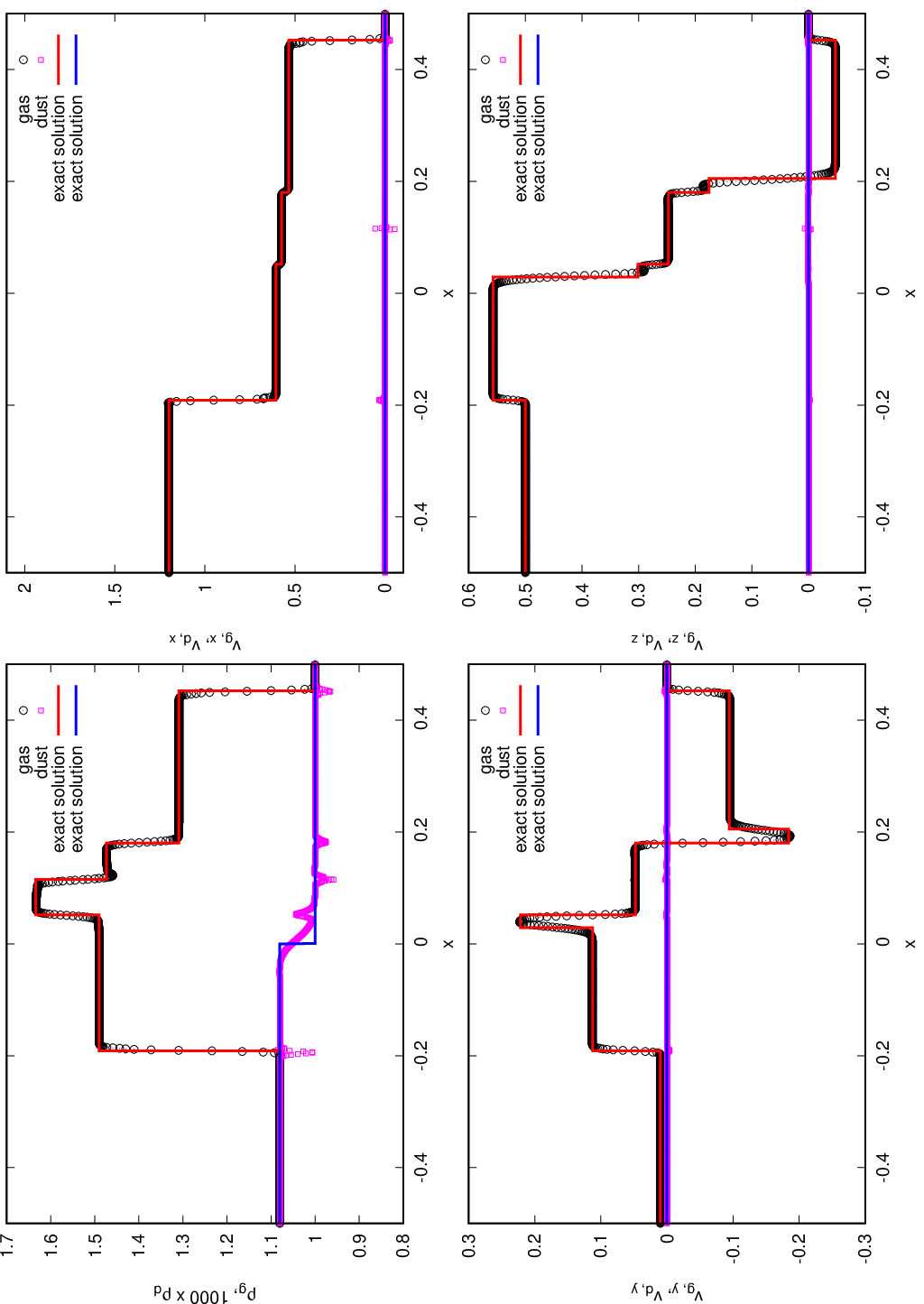}
\caption{
  Same as figure \ref{dusty_MHD_shock_strong} but in the weak coupling limit ($K=0$).
  Red and blue lines show the analytical solutions for the gas and dust, respectively.
}
\label{dusty_MHD_shock_weak}
\end{figure*}

Figure \ref{dusty_MHD_shock_weak}
shows the results of the dusty MHD shocktube problem in the weak coupling limit ($K=0$).
In the dust density profile, there are blips at the shocks and the discontinuities.
The velocity profiles also have the blips, though much less clear, at the same positions.
The existence of these blips suggest that the dust density around the shock may have
an error of $\lesssim 5$ percent in weak coupling limit.
Another important point is that
the discontinuity of the dust density at $x=0$ in the exact solution is smeared out in our numerical solution.
This is interpreted to be due to numerical dissipation introduced in our scheme.
Thus, our scheme tends to smoothen discontinuity of the dust density.
These two numerical artifacts should be noted in use of our scheme.

\subsection{gravitational collapse of the dusty magnetized cloud core}
The final test problem in this paper is
gravitational collapse of the magnetized cloud core.

The initial condition of the cloud core is as follows.
We adopt the density-enhanced Bonnor-Ebert sphere surrounded by medium with a steep density
profile of $\rho \propto r^{-4}$ as the initial density profile, which given by
\begin{eqnarray}
  \rho(r)=\begin{cases}
  \rho_0 \xi_{\rm BE}(r/a) ~{\rm for} ~ r < R_c \\
  \rho_0 \xi_{\rm BE}(R_c/a)(\frac{r}{R_c})^{-4} ~{\rm for} ~ R_c < r < 5 R_c,
  \end{cases}
\end{eqnarray}
and
\begin{eqnarray}
  a=c_{\rm s, iso} \left( \frac{f}{4 \pi G \rho_0} \right)^{1/2},
\end{eqnarray}
where $\xi_{\rm BE}$ is a non-dimensional density profile of the
critical Bonnor-Ebert sphere,
$f$ is a numerical factor related to the strength of the gravity,
and $R_c=6.45 a$ is the radius of the cloud core.
Specifically, $f=1$ corresponds to the critical Bonnor-Ebert sphere, and
the core with $f>1$ is gravitationally unstable.
In this study, we adopt $\rho_0=7.3\times 10^{-18} \gcm$,
$\rho_0/\rho(R_c)=14$, and $f=2.1$. Then, the radius of the core
is $R_c=4.8\times 10^3$ AU and the enclosed mass within $R_c$ is $M_c=1 \msun$.
The parameter $\alpha_{\rm therm}$ ($\equiv E_{\rm therm}/E_{\rm grav}$) is $0.4$, where
$E_{\rm therm}$ and $E_{\rm grav}$ are the thermal and gravitational energies of the central core
(without surrounding medium), respectively.
We adopt an angular velocity 
profile of $\Omega(d)=\Omega_0/(\exp(10(d/(1.5 R_c))-1)+1)$
with $d=\sqrt{x^2+y^2}$ and  $\Omega_0=2.3\times 10^{-13} {\rm s^{-1}}$.
With this formula $\Omega(d)$ is almost constant for $d<1.5 R_c$ and rapidly decreases for $d > 1.5 R_c$.
The ratio of the rotational to gravitational energies $\beta_{\rm rot}$ within
the core is $\beta_{\rm rot}$ ($\equiv E_{\rm rot}/E_{\rm grav}$) $=0.03$,
where $E_{\rm rot}$ is the rotational energy of the core.
We adopt a constant magnetic field $(B_x,B_y,B_z)=(0,0,83 \mu G)$.
The mass-to-flux ratio of the core is  $\mu/\mu_{\rm crit}=3$.
We resolve 1 $\msun$ with $3\times 10^6$ SPH particles. 
We adopt the dust density profile of $\rho_d(r)=f_{dg} \rho_g(r)/(\exp(10(r/(1.5 R_c))-1)+1)$
where $f_{dg}=10^{-2}$ is the dust-to-gas mass ratio.
The dust density profile has the same shape with the gas density profile in $r\lesssim 1.5 R_c$ but
is truncated at $r \geq 1.5 R_c$ to prevent  artificial dust accretion from the outer medium
in the simulation.

We assume the uniform dust size in each simulation run, applying 4 values of $a_d=1~\mum, 10~\mum, 100~\mum,$ and $ 1~ {\rm mm}$.

The dust friction law is given by \citep[][]{2012MNRAS.420.2365L}
\begin{eqnarray}
  K=   \frac{4}{3} \pi \rho_g a_d^2 v_{\rm therm} n_d \sqrt{1+\frac{9 \pi}{128} (\frac{\dv}{c_s})^2}
\end{eqnarray}
for the Epstein regime ($a_d<9/4 \lambda_{\rm mfp}$)
where $\lambda_{\rm mfp}=m_g/(\sigma_{\rm mol} \rho_g)$ is the mean free path
and $\sigma_{\rm mol}=2 \times 10^{-15} {\rm cm^2}$ is the collisional cross section of the gas molecule.

For the Stokes regime ($a_d >9/4 \lambda_{mfp}$), the parameter $K$ is given by
\begin{eqnarray}
  K= \frac{1}{2} C_d \pi a_d^2 \rho_g |\dv| n_d,
\end{eqnarray}
where
\begin{eqnarray}
  C_d= \begin{cases}
    0.44  (Re>800),\\
    24 Re^{-0.6}  (800>Re>1),\\
    24 Re^{-1} (1>Re).\\
      \end{cases}
\end{eqnarray}

Here, $Re=2 a_d |\dv|/\nu$ is the Reynolds number,
$\nu=5 \sqrt{\pi} \mu m_g/(64 \rho_g \sigma_{\rm mol}) c_s$
is the kinetic viscosity,
$v_{\rm therm}=\sqrt{8/\pi} c_s$ is the thermal velocity,
$n_d=\rho_d/m_d$ is the number density of the dust,
$m_d=\frac{4}{3} \pi \rho_{mat} a_d^3$ is the mass of the dust,
and $\rho_{mat}=2 \gcm$ is the internal density of the dust.
We include Stokes regime just for the consistency to the previous studies.
However, with the dust size adopted in this test, the entire region is in Epstein drag regime.

Our numerical simulations solve non-ideal MHD equations, including the Ohmic and ambipolar
diffusion but ignoring the  Hall effect.
The equations are the same as in our previous study \citep{2020ApJ...896..158T}.
For the resistivity model, we adopt the resistivity table with $a_d=0.035 \mum$ of
\citet{2020ApJ...896..158T}. Thus, the dust size for dust dynamics and that for the resistivity table
are not consistent with each other in our present simulations.

\begin{figure*}
  \includegraphics[clip,trim=0mm 0mm 0mm 0mm,width=150mm,,angle=-90]{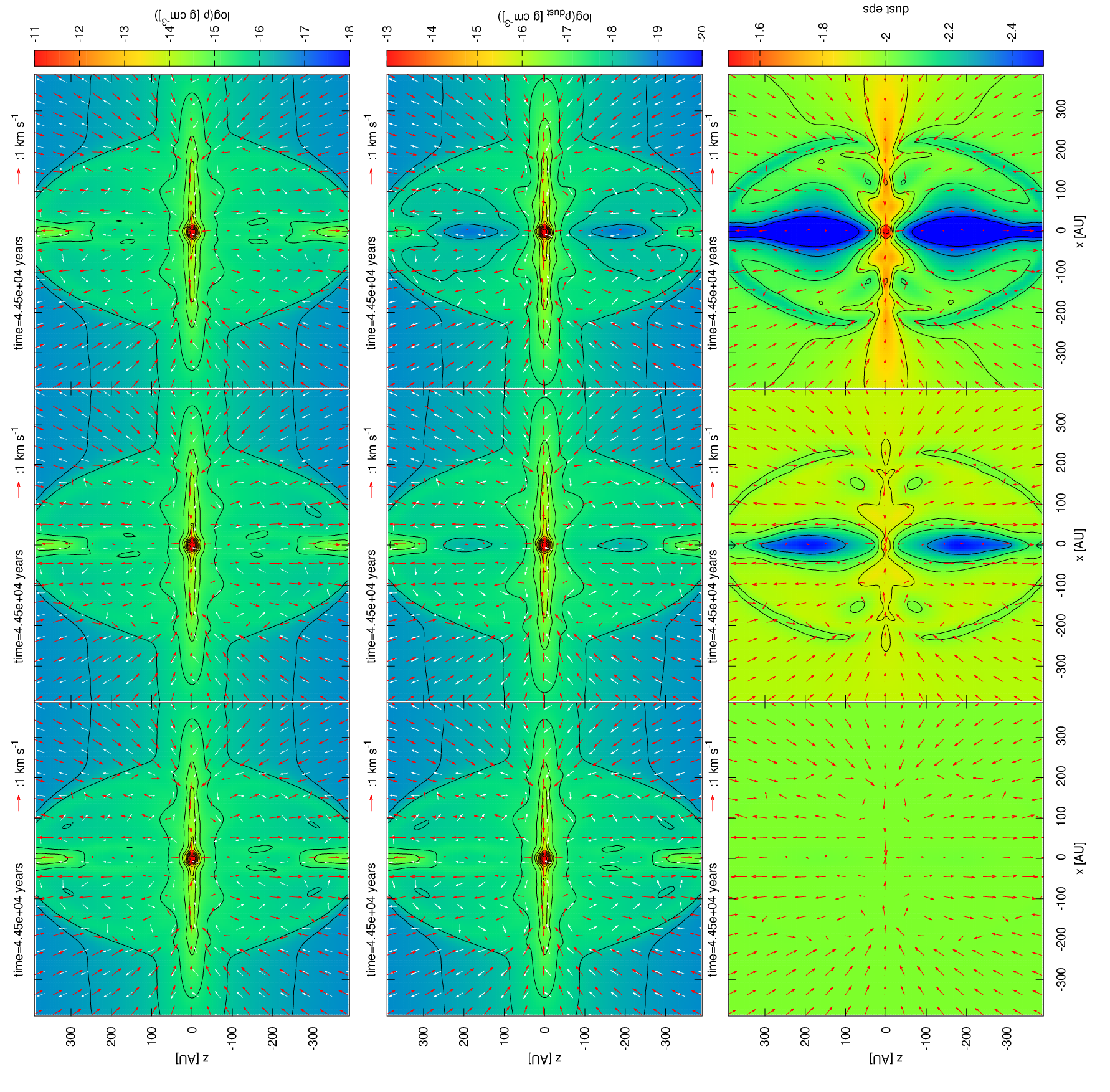}
\caption{
  Two dimensional maps of gas (top), dust density (middle), and $\epsilon$ (bottom).
  The box size is 800 AU. The elapsed time is $t=4.45 \times 10^4$ yr.
  Red arrows and white arrows show the velocity field and direction of the magnetic field, respectively.
  The results with $a_d=1 \mum$ (left), $a_d=100 \mum$ (middle), $a_d=1 {\rm mm}$ (right) is shown.
}
\label{dust_2D_map}
\end{figure*}

Figure \ref{dust_2D_map} shows the result of our simulation;
the gas (top panels), dust density (middle panels) and $\epsilon$ (bottom panels)
which roughly corresponds to the dust-to-gas mass ratio
for $a_d=1 \mum, 100 \mum, $ and $ 1 {\rm mm}$ at the end of the simulations ($t=4.45 \times 10^4$ yr)  are all plotted on the  $x$-$z$ plane.
For the micron-sized dusts ($a_d=1 \mum$), the dusts and gas 
density maps are identical to each other (top-left and middle-left panels) and they are completely coupled.
Thus, the dust-to-gas mass ratio $\epsilon$ is constant as expected (top-bottom panel).
Note that, if a change of the dust-to-gas mass ratio occurred in the outflow with $a_d=1 \mum$,
it would be a numerical artifact because
the approximation of equation (\ref{assumption11}) to (\ref{assumption13}) could break down in the outflow.

Note also that the constant dust-to-gas mass ratio is realized because of very short stopping time stem from
large density in the collapsing cloud core. In contrast, in molecular cloud, the dusts with size of $a_d\sim 1 \mum$ decouples from gas,
which causes a fluctuation of dust-to-gas mass ratio \citep{2017MNRAS.471L..52T}.

\begin{figure*}
  \includegraphics[clip,trim=0mm 0mm 0mm 0mm,width=200mm,,angle=-90]{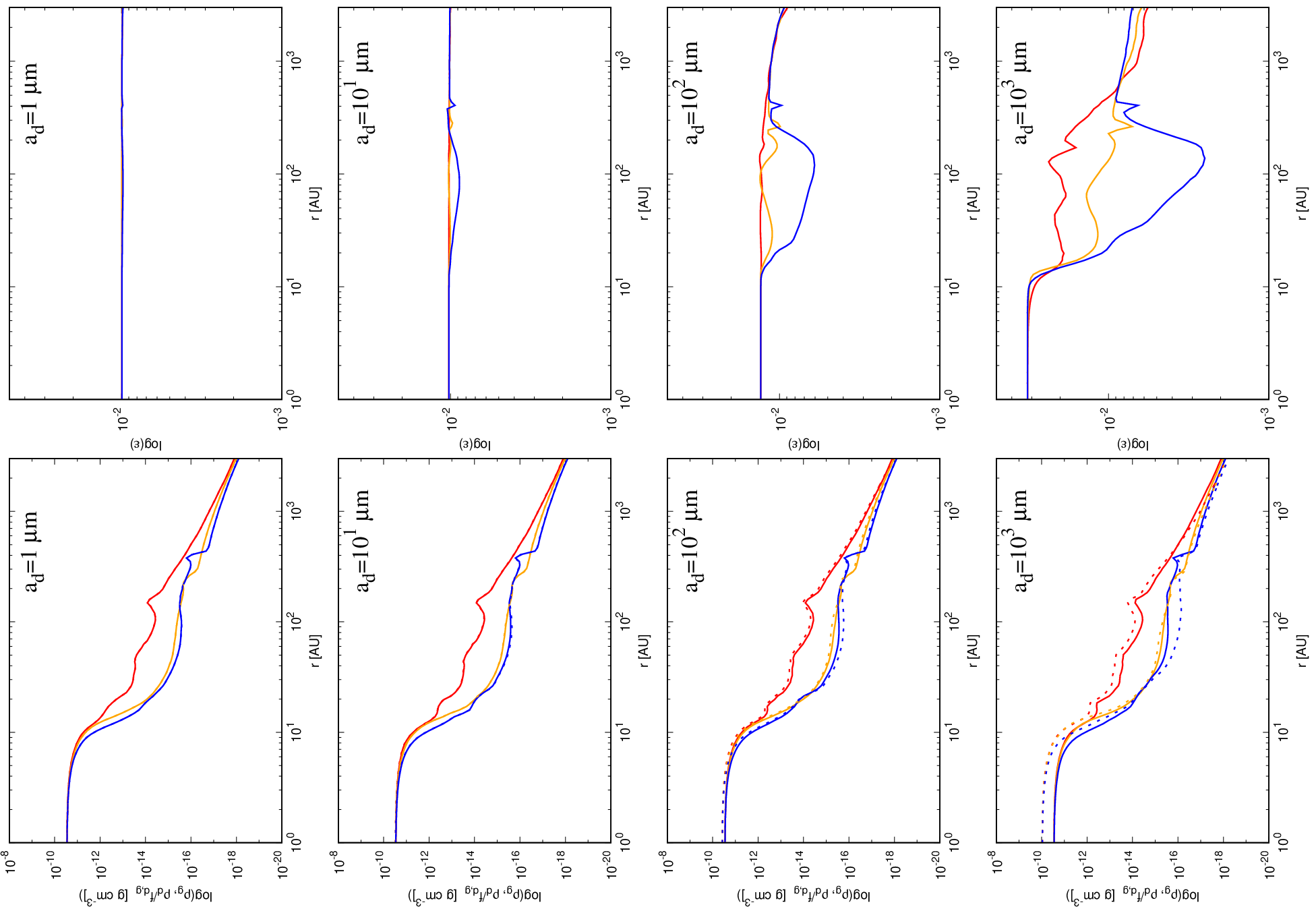}
\caption{
Left panels show the radial profiles of $\rho_g$ (solid lines) and $\rho_d/f_{d,g}=100 \rho_d$ (dashed lines)
from $a_d=1 \mum$ (top panel) to $a_d=1 {\rm mm}$ (bottom panel).
The elapsed time is $t=4.45 \times 10^4$ yr.
Right panels shows the profile of $\epsilon$.
Red, orange, and blue lines show the profiles along the $x$ axis,
the axis tilted at  $45^\circ$ from the $x$ axis, and $z$ axis, respectively.
}
\label{dust_1D_prof}
\end{figure*}

  Here, we have omitted the results with $a_d=10 \mum$ in figure \ref{dust_2D_map}
  to reduce the size of the figure; the results with $a_d=10 \mum$ are very similar to 
  the results with $a_d=1 \mum$ (see figure \ref{dust_1D_prof}).
  Although we did not execute the simulation with $a_d=0.1 \mum$ which is discussed in \S 2,
  they are expected to be the same as the results with $a_d=1 \mum$ because their stopping time is even smaller than that of $a_d=1 \mum$.

In the results with $a_d = 100 \mum$,  deviation from the constant dust-to-gas mass ratio is observed.
Specifically, at the midplane of the pseudo-disk (in the vicinity of the  x-axis),
$\epsilon$ is slightly enhanced.
By contrast, $\epsilon$ clearly decreases in the outflow.
These imply that the pseudo-disk tends to be dust rich and the outflow
tends to be dust poor as the dust size increases.
In the results with $a_d = 1 {\rm mm}$, the deviation is even more prominent.
Furthermore, the central first-core also has large $\epsilon$, suggesting that the
dust-rich first-core, and hence dust-rich circumstellar disk, forms
if the dust size has already grown to a size of  $a_d \gtrsim 1 {\rm mm}$ in the envelope.

Figure \ref{dust_1D_prof} shows the obtained radial profiles of $\rho_g$ 
and $\rho_d/f_{d,g}=100 \rho_d$ 
for a range of $a_d$ from $a_d=1 \mum$ to $a_d=1 {\rm mm}$,
as well as the profiles of $\epsilon$
along $x$ axis,  axis tilted at  $45^o$ from the $x$ axis, and $z$ axis, respectively.

With the micron-sized dust ($a_d=1 \mum$ to $a_d=10 \mum$)
the density profiles of dust and gas are indistinguishable from each other
and $\epsilon$ is almost constant radially.
With a dust size of $a_d=100 \mum$,
they are found to be slightly different from each other;
the dust density is slightly larger in the inner region of $<10 $ AU and at midplane (red lines in figure \ref{dust_1D_prof}),
implying that the dust density is enhanced,
whereas it slightly decreases at $z=50$ AU (blue lines)  which corresponds to the region inside of the outflow.
The dust-to-gas mass ratio becomes $\sim 0.006$ at the local minimum.
With dust size of $a_d=1 {\rm mm}$, the difference between gas and dust density profile is more prominent.
The dust-to-gas mass ratio is $\sim 0.03$  in the central region and
decreases to $\sim 0.002$ in the outflow.

  Note that the approximation of equation (\ref{assumption11}) to (\ref{assumption13})  is also valid for $1$ mm dusts
  because the approximation tends to get better as the dust size increases
  (this is a physical consequence of the decrease in the total charge of the dust).

  Finally, we explain the relationship between this test problem and the discussion in \S \ref{governing_equation}.
  In \S \ref{governing_equation}, we investigate the condition which justifies the approximation of equation (\ref{assumption11}) to (\ref{assumption13})
  using the analytical model for $a_d\ge 10 \mum$ and using the chemial network calculation for $a_d \le 1 \mum$.
  Then, we used the approximation for this test problem.
  Therefore, we use the analytic model in \S \ref{governing_equation} to justify the simulation results with $ad=10, 100, 1000 \mum$ 
  and use the chemical network to justify the simulation results with $a_d=1 \mum$ of this subsection.


\section{Summary and Future work}
In this paper, we develop a simulation scheme for dust-gas mixture in the magnetized medium,
carefully assessing the validity of the underlying assumptions for our scheme.

In \S 2, we investigated the condition in which the electro-magnetic force on the dust is negligible
and the approximation of equations (\ref{assumption11}) to (\ref{assumption13}),
which is core concept for our scheme is valid.
In \S 3, we presented the SPH discretization of our scheme,
where we newly introduced a numerical
diffusion term for $\epsilon$, which greatly stabilizes the scheme.

Our numerical scheme was applied to three test problems in \S 4.
We showed our scheme was capable of treating the dusty wave problem from
the strong to weak coupling limit, and
found that the results were consistent with the previous studies.
Furthermore, our scheme was also found to be capable of  treating MHD shock-tube problem
in any coupling case.
Numerical errors near the shock or discontinuity was less than a few \% even in the weak coupling limit
and was negligible in the strong coupling limit.
Our scheme was successfully applied to the gravitational collapse of magnetized dusty
cloud cores. The results are consistent with the previous studies \citep{2020A&A...641A.112L}.

In subsequent study, we will investigate the dust growth in the early evolution phase of
the circumstellar disk which is expected to bring new insights to this field.

\section*{Acknowledgments}
We thank Dr. Kazunari Iwasaki, Dr. Kengo Tomida,
and Dr. Satoshi Okuzumi for fruitful discussions.
We also thank anonymous referee for helpful comments.
The computations were performed on the Cray XC40/XC50 
system at CfCA of NAOJ and Cray CS400 at institute for information management and communication
of Kyoto University.
This work is supported by JSPS KAKENHI 
grant number 18H05437, 18K13581, 18K03703.

\appendix
\section{comparison between chemical calculation and analytical model}
In this appendix, we compare the chemical network calculation which is used for the calculation of $a_d=0.1 \mum,~ 1 \mum$ and
the analytical calculation of \citet{2009ApJ...698.1122O} which is used for the
calculation of $a_d=10 \mum,~ 100 \mum$ in figures \ref{relative_velocity_among_dust} to \ref{Jg_Jd}.

\begin{figure*}
  \includegraphics[clip,trim=0mm 0mm 0mm 0mm,width=35mm,,angle=-90]{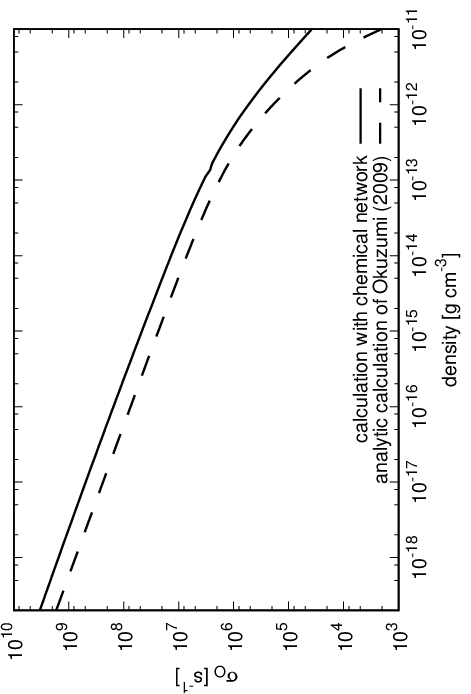}
  \includegraphics[clip,trim=0mm 0mm 0mm 0mm,width=35mm,,angle=-90]{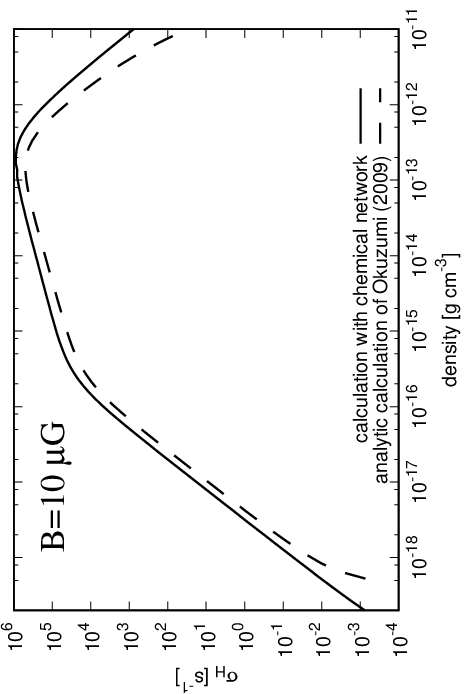}
  \includegraphics[clip,trim=0mm 0mm 0mm 0mm,width=35mm,,angle=-90]{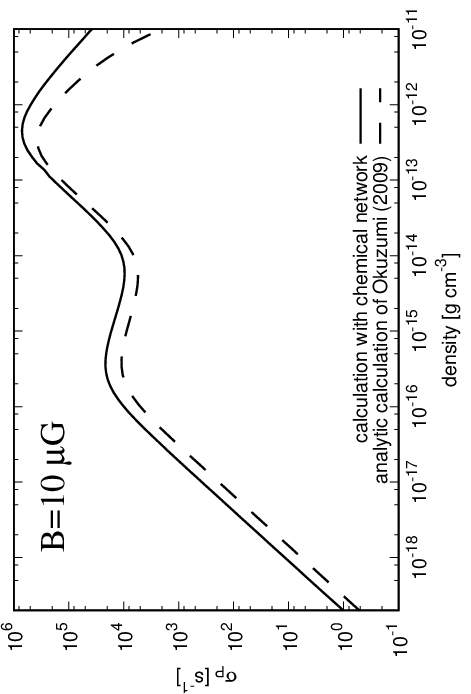}
  \caption{
    $\sigma_O$ (left), $\sigma_H$ (middle), and $\sigma_P$ (right) as a function of density. The solid and dashed lines show the
    results of chemical calculation and analytical calculation, respectively.
    $\sigma_H$ and $\sigma_P$ depend on the magnetic field strength and we fix $|\magB|=10 \mu G$ for them.
}
\label{comp_chemistory_analytic}
\end{figure*}

Figure \ref{comp_chemistory_analytic} shows
$\sigma_O$, $\sigma_H$, and $\sigma_P$ calculated from the chemical network calculation and analytical calculation for $a_d=1 \mum$.
The magnetic field strength for $\sigma_H$ and $\sigma_P$ is fixed to be $|\magB|=10 {\rm \mu G}$.
We checked the $\sigma_H$ and $\sigma_P$ with different magnetic field strength and our conclusion was confirmed
to be the same.

The figure shows that the difference of the conductivities between the two models is at most about one order of magnitude and shows
that the models are mutually consistent. Therefore we conclude that the analytical calculation is applicable
to evaluate the abundances of the charged species with large dusts of $a_d \gtrsim 1 \mum$.

\section{Lorentz transformation}
In this appendix, we present the Lorentz transformation for convenience of readers.
The Lorentz transformation between the laboratory frame and the neutral rest frame is given by
\begin{eqnarray}
  \eleE'=(1-\gamma)\frac{(\vel \cdot \eleE) \vel}{\vel^2}+\gamma (\eleE+\frac{\vel \times \magB}{c}), \\
  \magB'=(1-\gamma)\frac{(\vel \cdot \magB) \vel}{\vel^2}+\gamma (\magB-\frac{\vel \times \eleE}{c}), \\
  \tau'_{[g, d]}=\gamma(\tau_{[g, d]}- \frac{\vel \cdot \cul_{[g,d]}}{c^2}), \\
  \cul'_{[g, d]}=\cul_{[g, d]}+(\gamma-1)\frac{(\vel \cdot \cul_{[g, d]})\vel}{\vel^2}- \gamma \tau_{[g, d]} \vel,
\end{eqnarray}
where $\gamma=(1-\vel^2/c^2)^{-1/2}$ is the Lorentz factor.
By assuming $\gamma=1$ and $|\vel \times \eleE|/(c |\magB|) \sim v^2/c^2 \ll 1$,
we obtain
\begin{eqnarray}
  \eleE'&=&\eleE+\frac{\vel \times \magB}{c}, \\
  \magB'&=&\magB, \\
  \tau'_{[g, d]}&=&\tau_{[g, d]}- \frac{\vel \cdot \cul_{[g,d]}}{c^2}, \\
  \cul'_{[g, d]}&=&\cul_{[g, d]}+ \tau_{[g, d]} \vel.
\end{eqnarray}
These relations are used in \S \ref{sec_assumption}.
The relation of $|\vel \times \eleE|/(c |\magB|) \sim v^2/c^2 \ll 1$ in the case of non-ideal MHD
is deduced from the Ohm's law as
$|\vel \times \magB|/(c \eleE) \sim |\vel \times \magB|/(\eta \nabla \times B) \sim v L/\eta \sim Re_m$
where $Re_m$ is the magnetic Reynolds number and hence
$|\eleE|/|\magB| \sim  (v/c) Re_m^{-1} \sim (v/c)$ which is justfied when $Re_m \gtrsim 1$.

\bibliography{article}

\end{document}